\documentclass[12pt,twoside,fleqn]{article}
\usepackage{color,colortbl,graphicx,setspace,multirow,enumitem,float}
\usepackage[utf8]{inputenc}

\usepackage{geometry}
\setitemize{labelindent=1cm,labelsep=0.25cm,leftmargin=1cm}

\usepackage[hang,flushmargin]{footmisc} 
\usepackage[hang]{footmisc}
\usepackage{natbib} 
\usepackage{amsfonts,amsmath,amssymb} 
\usepackage{booktabs,threeparttable} 
\usepackage{placeins} 

\usepackage[dvipsnames]{xcolor}
\definecolor{nblue}{HTML}{000660}
\usepackage[colorlinks=true,urlcolor=nblue,linkcolor=nblue,citecolor=nblue]{hyperref}
\newcommand{\textsff}[1]{{\sffamily #1}}

\usepackage[title,titletoc]{appendix}

\usepackage{titlesec} 
\titleformat{\section}[block]{\bfseries\sffamily\large}{\thesection. }{0em}{\MakeUppercase} 
\titleformat{\subsection}[block]{\bfseries\sffamily\large}{\thesubsection. }{0em}{} 
\titleformat{\subsubsection}[block]{\large}{}{0em}{\sffamily\itshape} 

\let\oldsection\section
\renewcommand{\section}{\FloatBarrier\oldsection}
\let\oldsubsection\subsection
\renewcommand{\subsection}{\FloatBarrier\oldsubsection}



\newcommand{\lat}{\text{\texttt{l}}}
\newcommand{\obs}{\text{\texttt{o}}}
\newcommand{\res}{\text{\texttt{r}}}
\newcommand{\mis}{\text{\texttt{m}}}

\newcommand{\bdiag}{\text{bdiag}}
\newcommand{\diag}{\text{diag}}
\newcommand{\Pm}{\mathbb{P}}
\newcommand\given[1][]{\:#1\vert\:}

\usepackage{bm}
\usepackage{caption}
\usepackage{subcaption}

\makeatletter\let\p@subfigure\thefigure\makeatother
\newcommand\fnote[1]{\captionsetup{font=small}\caption*{#1}}

\usepackage{fancyhdr} 
\def\titletext{Large Bayesian VARs for Binary, Censored and Count Variables}
\title{\sffamily\huge{\textbf{\titletext}}}
\author{}
\date{}

\begin{document}

\maketitle
\vspace*{-4.5em} 
\normalsize
\begin{center}
\begin{minipage}{.32\textwidth}
  \large\centering Joshua C.C. Chan\\[0.25em]
  \normalsize\textit{Purdue University}
\end{minipage}
\begin{minipage}{.32\textwidth}
  \large\centering Davide Pettenuzzo\\[0.25em]
  \normalsize\textit{Brandeis University}
\end{minipage}
\begin{minipage}{.32\textwidth}
  \large\centering Michael Pfarrhofer\\[0.25em]
  \normalsize\textit{WU Vienna}
\end{minipage}\vspace*{0.5cm}

\begin{minipage}{.32\textwidth}
  \large\centering Aubrey Poon\\[0.25em]
  \normalsize\textit{University of Kent}
\end{minipage}
\begin{minipage}{.35\textwidth}
  \large\centering Dan Zhu\\[0.25em]
  \normalsize\textit{Monash University}
\end{minipage}

\vspace*{1em}

{\large September 2026}

\end{center}

\begin{center}
\begin{minipage}{0.85\textwidth}
\noindent\small\onehalfspacing

We extend the vector autoregression (VAR) to model binary, censored, count and unrestricted variables jointly, treating missing observations of any type in the same way. Each restricted observation is linked to a latent Gaussian variable, so that the standard machinery for VARs, such as shrinkage priors, stochastic volatility and outlier components, applies directly to the VAR for the latent variables. The latent variables have a truncated normal conditional distribution with a banded precision matrix, from which they are drawn, jointly or equation by equation, using Hamiltonian Monte Carlo, and the sampler scales well to high dimensions. A simulation study shows that treating restricted variables as continuous distorts the levels and scales of the system and worsens forecasts of the restricted outcomes in most designs. In an application to 25 US macro-financial variables, we estimate a business cycle indicator, the intensity of banking distress and shadow rates. We use the model for scenario analysis with restrictions on the recession probability, bank failures and interest rates at the effective lower bound.\\

\textsff{\textbf{JEL}}: C11, C32, C34, C35, C53, E32 \\
\textsff{\textbf{KEYWORDS}}: Bayesian VAR, binary and censored variables, count data, effective lower bound, shadow rate, bank failures, conditional forecasting
\end{minipage}
\end{center}

\singlespacing\vfill\noindent{\footnotesize\textit{Contact}: Michael Pfarrhofer (\href{mailto:michael.pfarrhofer@wu.ac.at}{michael.pfarrhofer@wu.ac.at}). This paper merges our previous work ``Modeling and Forecasting Count Data with Bayesian Vector Autoregressions'' (Pettenuzzo, Poon and Zhu; \texttt{SSRN 5285954}) and ``Large Bayesian VARs for Binary and Censored Variables'' (Chan and Pfarrhofer; \texttt{arXiv 2506.01422}). We acknowledge the use of Claude Code and Codex for drafting, editing and brainstorming. A previous version of the paper was reviewed by Refine, and its suggestions are incorporated in the current draft. All AI-assisted suggestions were carefully reviewed, and we take full responsibility for the final content and accuracy of this paper. Codes and replication files: \href{https://github.com/mpfarrho/protoc-var}{github.com/mpfarrho/protoc-var}.
}

\thispagestyle{empty}\renewcommand{\footnotelayout}{\setstretch{1.5}}\newpage
\doublespacing\normalsize\renewcommand{\thepage}{\arabic{page}}

\section{Introduction}

Since the seminal work of \citet{Sims80} and \citet{DLS84}, vector autoregressions (VARs) have become the workhorse models for macroeconomic forecasting and structural analysis. Because most macroeconomic variables are continuous, the literature has focused on real-valued time series. Many important macroeconomic and financial indicators, however, have restricted supports. Nominal interest rates are bounded below by their effective lower bound (ELB); recessions, financial crises and sovereign defaults are binary events; and bank failures, corporate bankruptcies and loan defaults are counts, as are many other outcomes in finance \citep{cohn2022count}. Standard VARs are ill-suited for such series, and the growing use of large datasets, which typically contain several of them, makes the issue more acute \citep{banbura2010large, koop2013forecasting}.

We develop an econometric framework that extends the VAR to a mix of binary, censored, count and unrestricted variables, with missing observations of any type handled in the same way. The approach is based on data augmentation: each restricted observation is linked to a real-valued latent variable, so that a binary variable records the sign of its latent variable as in a probit model, a censored variable is observed only above a threshold as in a tobit model, and a count variable records the interval between consecutive thresholds in which its latent variable falls, as in an ordered probit model. The thresholds for the counts are given by a link function from the Box--Cox family, which contains unit intervals and logarithmic thresholds as special cases. Given the latent variables, the machinery developed for standard VARs applies directly. In particular, the framework accommodates adaptive hierarchical shrinkage priors \citep{giannone2015prior}, which are essential in high dimensions, and stochastic volatility and outlier components, which have been shown to be empirically important for macroeconomic and financial variables \citep{cogley2005drifts, clark2011real, CCM19, chan23JE, carriero2021addressing}.\footnote{Bayesian data augmentation for binary and censored variables goes back to \citet{chib1992bayes} and \citet{AC93}; see \citet{anceschi2023bayesian} for a review.}

Sampling the latent variables is challenging in a high-dimensional dynamic setting for two reasons. First, the latent variables are subject to one inequality restriction for each restricted observation, so that their conditional distribution is a truncated normal in a dimension of the order of the sample size, for which accept-reject sampling almost always rejects. Second, they are serially dependent through the VAR, so that a Gibbs sampler that draws them one at a time mixes poorly. Building on \citet{chan2023high}, we show that the precision matrix of this truncated normal distribution is banded, and we sample from it, jointly or equation by equation, with the Zigzag Hamiltonian Monte Carlo (HMC) sampler of \citet{nishimura2020discontinuous} and \citet{nishimura2025zigzag}. A timing study shows that the sampler is feasible for VARs with as many as 30 variables, 750 observations and nine restricted variables.

In a Monte Carlo study, we compare the proposed model with a Gaussian VAR fitted directly to the observed outcomes across 24 designs. The proposed model forecasts the observed count variables better in every design, the binary and censored variables in most designs, and the continuous variables equally well. Treating the restricted variables as continuous distorts the intercepts and the innovation covariance matrix.

We then apply the framework to 25 monthly US macro-financial variables: the NBER recession indicator, the number of bank failures, six interest rates that are subject to censoring at the ELB, and 17 unrestricted variables, two of which are unavailable early in the sample. The latent variables are interpretable: a business cycle indicator underlying the NBER recession dates, the intensity of banking distress underlying the bank failure count, and shadow rates that summarize the stance of monetary policy at the ELB. Finally, we use the model for conditional forecasting and scenario analysis in which scenarios can be formulated for the restricted variables as well as for the continuous variables: a target for the recession probability on impact, the bank failure path of the Global Financial Crisis as the expected path, a tightening of financial conditions, the 2026 supervisory stress test scenarios, and oil price paths calibrated to the 2026 Iran war.

Our paper contributes to the literature on VARs with restricted variables. \citet{dueker2005dynamic} develops the qualitative VAR, which adds a single binary variable to a standard VAR and is used for forecasting and scenario analysis in \citet{dueker2006business}, \citet{dueker2010forecasting} and \citet{mccracken2022binary}. \citet{poon2024recessions} model several binary indicators jointly with a multinomial logit model. For censored variables, \citet{johannsen2021time} propose an unobserved components model for interest rates at the ELB, and \citet{carriero2023shadow} model interest rates as censored observations of latent variables within a VAR. Count time series have mostly been modeled one series at a time, by integer-valued autoregressions \citep{alosh1987first} or by latent Gaussian transformations \citep{jia2023latent}; see \citet{davis2021count} and \citet{fokianos2024multivariate} for reviews. Multivariate count models and copula models for mixed data introduce the dependence across series separately from the univariate margins \citep{HEINEN_2007,berry2020bayesian,zito2025dynamic}. Relative to this literature, our framework handles multiple variables of each type and missing values within one latent Gaussian VAR, and we develop a posterior sampler that scales to the dimensions of applied work.

The paper is organized as follows. Section~\ref{s:framework} introduces the VAR for binary, censored, count and unrestricted variables, as well as the posterior sampler for the latent variables. Section~\ref{s:simevidence} first documents the computational cost of the samplers across 24 designs and then reports the forecast accuracy of the proposed model relative to a Gaussian VAR fitted to the observed outcomes. Section~\ref{s:application} applies the framework to US macro-financial data, reports the estimates of the latent states and conducts the scenario analysis. Section~\ref{s:conclusion} concludes. 

\section{Econometric Framework} \label{s:framework}

This section introduces a VAR for a mix of binary, censored, count and unrestricted variables and discusses the normalizations that identify it. It then derives the conditional distribution of the latent variables, outlines the sampler that draws from it, and states the priors.

\subsection{VAR for Mixed Variables} \label{sec:VAR}

To set the stage, let $\widetilde{\bm{y}}_t = (\widetilde{\bm{y}}_{b,t}',\widetilde{\bm{y}}_{c,t}',\widetilde{\bm{y}}_{z,t}',\bm{y}_{u,t}')'$, for $t=1,\hdots,T,$ be an $n \times 1$ vector, with $n = n_b + n_{c} + n_{z} + n_u$, consisting of: 
\begin{itemize}
    \item $n_b$ binary variables $\widetilde{y}_{b,it} \in \{0,1\}$ in $\widetilde{\bm{y}}_{b,t} = (\widetilde{y}_{b,1t},\hdots,\widetilde{y}_{b,n_bt})'$;
    \item $n_{c}$ censored variables $\widetilde{y}_{c,it} \in [l_i,\infty)$ in $\widetilde{\bm{y}}_{c,t} = (\widetilde{y}_{c,1t},\hdots,\widetilde{y}_{c,n_{c}t})'$ with known thresholds $l_i$ (for ease of exposition we take $l_i=0$; the proposed framework can be easily adapted to cases with non-zero thresholds, or when observations are censored from above);
    \item $n_{z}$ count variables with support in the set of non-negative integers, $\widetilde{y}_{z,it} \in \mathbb{Z}_{\geq 0} $, in $\widetilde{\bm{y}}_{z,t} = (\widetilde{y}_{z,1t},\hdots,\widetilde{y}_{z,n_{z}t})'$;
    \item $n_u$ unrestricted (continuous) variables $y_{u,it} \in \mathbb{R}$ in $\bm{y}_{u,t} = (y_{u,1t},\hdots,y_{u,n_ut})'$.
\end{itemize}

Due to the restricted supports of the subvectors $\widetilde{\bm{y}}_{b,t}$, $\widetilde{\bm{y}}_{c,t}$ and $\widetilde{\bm{y}}_{z,t}$, it is inappropriate to model $\widetilde{\bm{y}}_t$ using a VAR with errors drawn from a continuous distribution. A standard approach to handling discrete and censored variables is via data augmentation: introduce latent variables $\bm{y}_{b,t}\in\mathbb{R}^{n_b}$, $\bm{y}_{c,t}\in\mathbb{R}^{n_c}$ and $\bm{y}_{z,t}\in\mathbb{R}^{n_z}$ that are linked to, respectively, the binary observations $\widetilde{\bm{y}}_{b,t}$, the censored observations $\widetilde{\bm{y}}_{c,t}$, and the count variables $\widetilde{\bm{y}}_{z,t}$ through 
\begin{align}
    \widetilde{y}_{b,it} &= \mathbb{I}(y_{b,it} > 0), \quad \text{for } i = 1,\hdots,n_b,\label{eq:linkbinary}\\
    \widetilde{y}_{c,it} &= y_{c,it}\mathbb{I}(y_{c,it} > 0) = \max(y_{c,it},0), \quad \text{for } i = 1,\hdots,n_c,\label{eq:linkcensored}\\
    \widetilde{y}_{z,it} &= \sum_{k=1}^{\infty}\mathbb{I}\left(y_{z,it} \geq g_i(k)\right), \quad \text{for } i = 1,\hdots,n_z,\label{eq:linkcount}  
\end{align}
where $\mathbb{I}(E)$ is an indicator function that takes the value 1 if the event $E$ is true and 0 otherwise, and $g_i$ is a strictly increasing function on the positive integers with $g_i(1) = 0$. The link functions \eqref{eq:linkbinary} and \eqref{eq:linkcensored}, respectively, give the standard probit and tobit models. Under \eqref{eq:linkcount}, the observed count is the number of thresholds $g_i(1) < g_i(2) < \cdots$ that lie below the latent variable: $\widetilde{y}_{z,it} = 0$ if $y_{z,it} < 0$, and $\widetilde{y}_{z,it} = k$ if $g_i(k) \leq y_{z,it} < g_i(k+1)$ for $k \geq 1$. Equation \eqref{eq:linkcount} is therefore an ordered probit model with fixed thresholds, extended to an unbounded support. We use the Box--Cox family, $g_i(k) = (k^{\psi_i} - 1)/\psi_i$, $\psi_i \in [0,1],$ with $g_i(k) = \log k$ for $\psi_i = 0$. For $\psi_i = 1$ the thresholds are the non-negative integers, each count corresponds to an interval of unit length, and \eqref{eq:linkcount} reduces to $\widetilde{y}_{z,it} = \max(\lceil y_{z,it}\rceil, 0)$ except at the integer values of $y_{z,it}$, which occur with probability zero; for $\psi_i = 0$ the count $k$ corresponds to the interval $[\log k, \log(k+1))$ of width $\log(1 + 1/k) \approx 1/k$, and \eqref{eq:linkcount} reduces to $\widetilde{y}_{z,it} = \lfloor \exp(y_{z,it}) \rfloor$. For every $\psi_i$ a zero count corresponds to the half-line $y_{z,it} < 0$.

The scale of $g_i$ is a normalization: multiplying $g_i$ and $y_{z,it}$ by the same positive constant leaves the distribution of the counts unchanged, in the same way that the unit error variance is a normalization in a probit model. The spacing of the thresholds, together with the Gaussian latent process introduced below, restricts the conditional distribution of the count, which is a discretized normal: $\Pm(\widetilde{y}_{z,it} = k) = \Phi((g_i(k+1) - \mu_{it})/\sigma_{it}) - \Phi((g_i(k) - \mu_{it})/\sigma_{it})$ for $k \geq 1$ and $\Pm(\widetilde{y}_{z,it} = 0) = \Phi(-\mu_{it}/\sigma_{it})$, where $\mu_{it}$ and $\sigma_{it}^2$ are the conditional mean and variance of $y_{z,it}$ and $\Phi$ is the standard normal distribution function. For $\psi_i = 1$, the thresholds are equally spaced. Away from the zero boundary, symmetry and dispersion depend on the latent mean's position relative to the threshold grid. When the latent standard deviation is sufficiently large relative to the unit spacing, these discretization effects are small, which gives an approximately symmetric distribution whose variance is approximately independent of the latent mean, holding the latent variance fixed. For $\psi_i = 0$ the distribution is right-skewed and its standard deviation is proportional to its mean for large counts. For every $\psi_i$ the zero cell collects all the mass below zero. The choice of $\psi_i$ affects the sampler only through the interval bounds of the count observations.

Poisson and negative binomial regressions are the standard tools for a single count series. In the Poisson distribution the conditional variance equals the conditional mean and in the negative binomial distribution it is a quadratic function of the mean, whereas under \eqref{eq:linkcount} the location and the scale of the conditional distribution are not tied to each other. Its shape is governed by one fixed parameter rather than by one threshold for each distinct count, as in an ordered probit model with free thresholds \citep[see also][]{kowal2020simultaneous}. Multivariate count models specify univariate margins for the counts and introduce the dependence across series separately, through a copula \citep{HEINEN_2007} or through shared latent factors \citep{berry2020bayesian}, as do copula models for mixed data \citep{zito2025dynamic}; in \eqref{eq:linkbinary} to \eqref{eq:linkcount} the margins and the dependence between all variables are instead implied jointly by the Gaussian VAR for the latent vector we discuss next, with a common lag structure and contemporaneous correlations.

Finally, we define $\bm{y}_t = (\bm{y}_{b,t}',\bm{y}_{c,t}',\bm{y}_{z,t}',\bm{y}_{u,t}')'$ which collects the latent variables associated with the binary, censored and count variables, as well as the observed continuous variables. This vector can be modeled as a VAR($P$) process:
\begin{equation}
    \bm{y}_t = \bm{A}' \bm{x}_t + \bm{u}_t, \quad \bm{u}_t = \bm{A}_0^{-1}\bm{O}_t\bm{H}_t^{1/2}\bm{\epsilon}_t,  \quad \bm{\epsilon}_t \sim \mathcal{N}(\bm{0}_n, \bm{I}_n),\label{eq:VAR}
\end{equation}
where $\bm{x}_t = (\bm{y}_{t-1}',\hdots,\bm{y}_{t-P}',1)'$ and $\bm{A} = (\bm{A}_1, \ldots, \bm{A}_P, \bm{a})'$ stacks the lag-specific $n\times n$ coefficient matrices $\bm{A}_p$ for $p = 1,\hdots,P,$ and $\bm{a}$ is a vector of intercepts. The regressor vector $\bm{x}_t$ can also include deterministic terms or exogenous variables, which enter the derivations below through the intercept term; we omit them to simplify the notation. The matrix $\bm{A}_0$ is unrestricted and invertible. The diagonal matrix $\bm{H}_t^{1/2} = \diag(\exp(h_{1t}/2),\hdots,\exp(h_{nt}/2))$ contains the standard deviations of the structural shocks, whose log-volatilities $h_{it}$ follow the processes specified in the next section. The outlier component $\bm{O}_t = \diag(o_{1t},\hdots,o_{nt})$ consists of scale factors that absorb large idiosyncratic shocks and outliers, as in \citet{carriero2021addressing}. Conditional on $\bm{H}_t$ and $\bm{O}_t$ the innovations are Gaussian, so that unconditionally they follow a scale mixture of normals. The reduced-form covariance matrix is $\mathrm{Var}(\bm{u}_t) = \bm{\Sigma}_t = \bm{A}_0^{-1}\bm{O}_t\bm{H}_t\bm{O}_t\bm{A}_0^{-1}{'}$.

To summarize, the proposed framework incorporates multivariate probit, tobit and ordered probit models to construct a VAR, which we call the ProToC-VAR for its probit, tobit and count components. It generalizes a variety of VARs previously considered in the literature. When $n_b = n_c = n_z = 0$, it reduces to the conventional VAR with continuous variables; when $n_c = n_z = 0$, it becomes a multivariate generalization of the Qual VAR of \citet{dueker2005dynamic} that contains a single binary variable; when $n_b = n_c = 0$, it reduces to a VAR for count and continuous variables; when $n_b = n_z = 0$, it becomes what has been termed a simple shadow rate VAR by \citet{carriero2023shadow} in an application to infer interest rates at the effective lower bound (ELB).

\subsection{Normalizations}\label{sec:norm}
There are two identification issues in this model. First, since $\bm{A}_0$ is unrestricted, the scale of the volatilities in $\bm{H}_t$ is not identified separately from the scale of $\bm{A}_0$. We follow \citet{chan2024large} and fix the unconditional mean of the log-volatilities at zero through the state equations $h_{it} = \phi_i h_{i,t-1} + u_{h,it}$, $u_{h,it}\sim\mathcal{N}(0,\omega_i^2)$, where $|\phi_i|<1$ and the initial condition is $h_{i0}\sim\mathcal{N}(0,\omega_i^2/(1-\phi_i^2))$. The presence of stochastic volatility identifies $\bm{A}_0$ up to permutations and sign changes of its rows. All objects reported in this paper, namely the reduced-form covariance matrices $\bm{\Sigma}_t$, the latent states, the forecasts and the generalized impulse responses, are invariant to this labeling. For diagnostic purposes, the stored draws of $\bm{A}_0$ are relabeled so that the diagonal elements are positive and the row permutation is closest to the identity.

Second, since the binary observations $\widetilde{\bm{y}}_{b,t}$ identify only the signs of the corresponding latent variables, the diagonal elements of $\bm{\Sigma}_t$ that correspond to the binary variables are not identified. In the spirit of \citet{dueker2005dynamic}, we normalize each draw of $\bm{A}_0$ so that the diagonal elements of the time-average innovation covariance matrix that correspond to the binary variables equal one. This matrix excludes the outlier component $\bm{O}_t$, $\bm{\bar{\Sigma}} = \sum_{t=1}^{T}\bm{A}_0^{-1}\bm{H}_t\bm{A}_0^{-1}{}' / T,$, so that the unit-scale restriction is a statement about normal times.

In each MCMC iteration, we sample $\bm{A}_0$, the log-volatilities and the outlier component from their unrestricted conditional posteriors. We then compute, for each binary variable $i$, the time-average innovation variance: $v_i = [\bm{\bar{\Sigma}}]_{ii} = \sum_{j=1}^{n}\left(\bm{A}_0^{-1}\right)_{ij}^2 \, \sum_{t=1}^{T}\exp(h_{jt})/T$, and form $\bm{A}_0^{\ast} = \bm{A}_0\bm{D}$, where $\bm{D} = \diag(d_1,\hdots,d_n)$ with $d_i = \sqrt{v_i}$ for the binary variables and $d_i = 1$ otherwise. Since $(\bm{A}_0^{\ast})^{-1} = \bm{D}^{-1}\bm{A}_0^{-1}$, the rescaled draw satisfies the normalization exactly, while the volatility paths $h_{jt}$ are unaffected. Rescaling $\bm{A}_0$ in this way is the change that a division of the latent variable $y_{b,it}$ by $\sqrt{v_i} > 0$ would imply for $\bm{A}_0$. Such a division preserves signs, so the likelihood of the observed binary data is unaffected by the scale that the normalization fixes. Only $\bm{A}_0$ is rescaled in this step. The latent states and the VAR coefficients keep their current values and are redrawn conditional on the normalized draw in the subsequent steps. The precision matrix $\bm{\Sigma}_t^{-1} = \bm{A}_0^{\ast}{}'\bm{H}_t^{-1}\bm{O}_t^{-2}\bm{A}_0^{\ast}$ used for sampling the latent states still includes the outlier component. We replace $\bm{A}_0$ by $\bm{A}_0^{\ast}$, so that all remaining MCMC steps condition on the normalized draw.

The rescaling is a deterministic step inserted between the conditional draws, as in \citet{dueker2005dynamic}. The resulting scheme is a draw-then-rescale device rather than a Gibbs sampler for the normalized model in the strict sense, but the normalized draws are the objects that the likelihood identifies, and the simulation study in Section~\ref{sec:sim} evaluates the parameter estimates on this normalized scale.\footnote{In a preliminary analysis, we consider several alternatives. Normalizing the stored draws ex post leaves the unidentified scale free to drift along the chain, which destabilizes the latent-state draws numerically. Fixing the diagonal elements of $\bm{A}_0$ that correspond to the binary variables through tight priors leaves the marginal variance of the latent index unpinned, which complicates the interpretation of its scale, and causes convergence problems in our experiments. Using $\bm{A}_0^{\ast}$ only in the step that draws the latent states, so that the parameter draws remain unrestricted, imposes the normalization only approximately and performs worse than the exact rescaling of every draw.}

\subsection{Sampling the Latent States}\label{subsec:latent}
It is useful to divide $\bm{y}_t$ into a subvector $\bm{y}_{\lat,t}$ of latent elements and a subvector $\bm{y}_{\obs,t}$ of observed elements (with subscripts \texttt{l} and \texttt{o}). Latent elements arise in two ways: from the data augmentation for the binary, censored and count variables, and from missing observations of any variable type, for example at the beginning or the end of the sample because of an unbalanced panel or a release delay. We partition $\bm{y}_t$ accordingly into three groups: the missing values $\bm{y}_{\mis,t}$ (labeled \texttt{m}, irrespective of variable type); the restricted values $\bm{y}_{\res,t}$ (labeled \texttt{r}), which are the latent variables underlying the observed binary and count variables and the censored observations; and the observed unrestricted values $\bm{y}_{\obs,t}$ (labeled \texttt{o}), which are the observed continuous variables and the uncensored observations of the censored variables. Let $n_t^{\mis}$, $n_t^{\res}$ and $n_t^{\obs}$ denote the numbers of elements in the three groups, and let $\bm{S}_{\mis,t}$, $\bm{S}_{\res,t}$ and $\bm{S}_{\obs,t}$ be the $n \times n_t^{\mis}$, $n \times n_t^{\res}$ and $n \times n_t^{\obs}$ selection matrices with $\bm{S}_{\mis,t}'\bm{y}_t = \bm{y}_{\mis,t}$, $\bm{S}_{\res,t}'\bm{y}_t = \bm{y}_{\res,t}$ and $\bm{S}_{\obs,t}'\bm{y}_t = \bm{y}_{\obs,t}$. The matrix $\bm{S}_t = [\bm{S}_{\mis,t}, \bm{S}_{\res,t}, \bm{S}_{\obs,t}]$ is an $n \times n$ permutation matrix, so that $\bm{S}_t\bm{S}_t' = \bm{I}_n$ and $\bm{S}_t'\bm{y}_t = (\bm{y}_{\mis,t}', \bm{y}_{\res,t}', \bm{y}_{\obs,t}')'$. Premultiplying the latter by $\bm{S}_t$ gives
\begin{equation}
    \bm{y}_t = \bm{S}_{\mis,t} \bm{y}_{\mis,t} + \bm{S}_{\res,t} \bm{y}_{\res,t} + \bm{S}_{\obs,t} \bm{y}_{\obs,t} = \bm{S}_{\lat,t} \bm{y}_{\lat,t} + \bm{S}_{\obs,t} \bm{y}_{\obs,t},\label{eq:decompY}
\end{equation}
where $\bm{S}_{\lat,t} = [\bm{S}_{\mis,t}, \bm{S}_{\res,t}]$ and $\bm{y}_{\lat,t} = (\bm{y}_{\mis,t}', \bm{y}_{\res,t}')'$ has $n_t^{\lat} = n_t^{\mis} + n_t^{\res}$ elements. The total number of latent elements is $N^{\lat} = \sum_{t=-P+1}^{T} n_t^{\lat}$, where the presample values $t = -P+1, \ldots, 0$ are treated in the same way as the in-sample values.

The next step is to derive the joint density of $\bm{y} = (\bm{y}_{-P+1}',\ldots,\bm{y}_{0}',\bm{y}_{1}',\ldots,\bm{y}_{T}')'$, which is of dimension $(T+P)n\times 1$. Let $\bm{y}_{\lat}$ and $\bm{y}_{\obs}$ denote, respectively, the vectors of all latent variables and observed data. Furthermore, let $\bdiag(\cdot)$ represent the operator that constructs a block diagonal matrix from its inputs. We define the selection matrices $\bm{S}_{\lat} = \bdiag(\bm{S}_{\lat,(-P+1)},\hdots,\bm{S}_{\lat,T})$ and $\bm{S}_{\obs} = \bdiag(\bm{S}_{\obs,(-P+1)},\hdots,\bm{S}_{\obs,T})$, so that $\bm{S}_{\lat}'\bm{y} = \bm{y}_{\lat}$ and $\bm{S}_{\obs}'\bm{y} = \bm{y}_{\obs}$. Stacking \eqref{eq:decompY} over $t = -P+1, \ldots, T$ gives
\begin{equation} \label{eq:y_partition}
 \bm{y} = \bm{S}_{\lat} \bm{y}_{\lat} + \bm{S}_{\obs} \bm{y}_{\obs}.
\end{equation}
Following \citet{chan2023high}, we encode the VAR dynamics in the $Tn \times (T+P)n$ matrix
\begin{equation*}
    \widetilde{\bm{M}} = \begin{bmatrix}
        -\bm{A}_P & \cdots & -\bm{A}_1 & \bm{I}_n & \bm{0}_{n\times n} & \cdots & \cdots & \bm{0}_{n\times n}\\
        \bm{0}_{n\times n} & -\bm{A}_P & \cdots & -\bm{A}_1 & \bm{I}_n & \bm{0}_{n\times n} & \cdots & \bm{0}_{n\times n}\\
        \vdots & \ddots & \ddots & & \ddots & \ddots & \ddots & \vdots\\
        \bm{0}_{n\times n} & \cdots & \bm{0}_{n\times n} & -\bm{A}_P & \cdots & -\bm{A}_1 & \bm{I}_n & \bm{0}_{n\times n}\\
        \bm{0}_{n\times n} & \cdots & \cdots & \bm{0}_{n\times n} & -\bm{A}_P & \cdots & -\bm{A}_1 & \bm{I}_n
    \end{bmatrix}.
\end{equation*}
Moreover, $\widetilde{\bm{\Omega}} = \bdiag(\bm{\Sigma}_1,\hdots,\bm{\Sigma}_T)$ is the $Tn \times Tn$ error covariance matrix, $\widetilde{\bm{m}} = (\bm{\iota}_T \otimes \bm{a})$ is the $Tn$-vector of intercepts, $\bm{\iota}_T$ is a $T$-vector of ones, and $\otimes$ is the Kronecker product. The presample vector $(\bm{y}_{-P+1}', \ldots, \bm{y}_{0}')'$ has prior mean $\bm{m}_0 = \bm{0}_{nP}$ and prior covariance $\bm{\Omega}_0 = \bm{I}_{nP}$. We accordingly define $\bm{M}_{0} = [\bm{I}_{nP}, \bm{0}_{nP\times Tn}]$ to obtain $\bm{M} = [\bm{M}_{0}', \widetilde{\bm{M}}']'$, $\bm{\Omega} = \bdiag(\bm{\Omega}_0,\widetilde{\bm{\Omega}})$, $\bm{m} = (\bm{m}_0',\widetilde{\bm{m}}')'$, so that \eqref{eq:VAR} and the prior on the initial conditions can be written as
\begin{equation}
    \bm{M}\bm{y} = \bm{m} + \bm{e}, \quad \bm{e}\sim\mathcal{N}(\bm{0}_{(T+P)n},\bm{\Omega}).\label{eq:static}
\end{equation}
Substituting \eqref{eq:y_partition} into \eqref{eq:static}, we have $\bm{M}\bm{S}_{\lat} \bm{y}_{\lat} + \bm{M}\bm{S}_{\obs} \bm{y}_{\obs} = \bm{m} + \bm{e}$. We now derive the distribution of the latent vector $\bm{y}_{\lat}$ conditional on the observed vector $\bm{y}_{\obs}$ and on the parameters. Defining $\bm{G}_{\lat} = \bm{M}\bm{S}_{\lat}$ and $\bm{G}_{\obs} = \bm{M}\bm{S}_{\obs}$, the exponent of the joint density implied by \eqref{eq:static} is $-\frac{1}{2} (\bm{G}_{\lat}\bm{y}_{\lat} + \bm{G}_{\obs}\bm{y}_{\obs} - \bm{m})'\bm{\Omega}^{-1}(\bm{G}_{\lat}\bm{y}_{\lat} + \bm{G}_{\obs}\bm{y}_{\obs} - \bm{m})$, which is quadratic in $\bm{y}_{\lat}$. Completing the square, as in \citet{chan2023high}, gives $\bm{y}_{\lat}\given(\bm{y}_{\obs},\bullet)\sim\mathcal{N}(\bm{\mu}_{\lat},\bm{V}_{\lat})$ with
\begin{equation}
    \bm{V}_{\lat} = (\bm{G}_{\lat}'\bm{\Omega}^{-1}\bm{G}_{\lat})^{-1}, \quad \bm{\mu}_{\lat} = \bm{V}_{\lat}\bm{G}_{\lat}'\bm{\Omega}^{-1}(\bm{m} - \bm{G}_{\obs}\bm{y}_{\obs}).\label{eq:poststates}
\end{equation}
Since $\bm{M}$ is block banded with bandwidth $P$ and $\bm{\Omega}$ is block diagonal, the precision matrix $\bm{V}_{\lat}^{-1} = \bm{G}_{\lat}'\bm{\Omega}^{-1}\bm{G}_{\lat}$ is banded when the latent elements are ordered chronologically. All computations below exploit this structure through sparse and band matrix routines.

Conditioning in addition on the observed binary, censored and count variables $\widetilde{\bm{y}}_b$, $\widetilde{\bm{y}}_c$ and $\widetilde{\bm{y}}_z$, we obtain $p(\bm{y}_{\lat}\given\widetilde{\bm{y}}_b,\widetilde{\bm{y}}_c,\widetilde{\bm{y}}_z,\bm{y}_{\obs},\bullet)$, which is a truncated multivariate normal distribution:
\begin{equation}
\bm{y}_{\lat}\given(\widetilde{\bm{y}}_b,\widetilde{\bm{y}}_c,\widetilde{\bm{y}}_z,\bm{y}_{\obs},\bullet) \sim \mathcal{N}(\bm{\mu}_{\lat},\bm{V}_{\lat}), \quad \text{subject to} \quad \bm{l} \leq \bm{y}_{\lat} \leq \bm{u}.\label{eq:truncGauss}
\end{equation}
The lower bound $\bm{l} = (l_1,\hdots,l_{N^{\lat}})'$ and upper bound $\bm{u} = (u_1,\hdots,u_{N^{\lat}})'$ are determined by the measurements in \eqref{eq:linkbinary} to \eqref{eq:linkcount}. In case an observation is missing, irrespective of the variable type, we have $(l_j, u_j) = (-\infty,\infty)$. The remaining cases are as follows:
\begin{itemize}
    \item Binary variables: $\widetilde{y}_{b,it} = 1$ gives $(l_j, u_j) = (0, \infty)$ and $\widetilde{y}_{b,it} = 0$ gives $(l_j, u_j) = (-\infty, 0)$;
    \item Censored variables: $\widetilde{y}_{c,it} = 0$ gives $(l_j, u_j) = (-\infty, 0)$;
    \item Count variables: $\widetilde{y}_{z,it} = 0$ gives $(l_j, u_j) = (-\infty, 0)$ and $\widetilde{y}_{z,it} = k > 0$ gives $(l_j, u_j) = (g_i(k), g_i(k+1))$, which are the unit intervals $(k-1, k)$ for $\psi_i = 1$ and the intervals $(\log k, \log(k+1))$ for $\psi_i = 0$.
\end{itemize}

\paragraph{Blocked updates.} Drawing all latent elements jointly from \eqref{eq:truncGauss} is possible but slow when $n$ and $T$ are large. Alternatively, the latent elements can be updated equation by equation, conditional on the current values of all other elements of $\bm{y}$. For variable $i$, let $\bm{S}^{(i)}_{\lat}$ be the selection matrix that picks the latent elements of variable $i$ out of $\bm{y}$ and $\bm{S}^{(i)}_{\obs}$ the selection matrix that picks all remaining elements, latent or observed, so that $\bm{y} = \bm{S}^{(i)}_{\lat}\bm{y}^{(i)}_{\lat} + \bm{S}^{(i)}_{\obs}\bm{y}^{(i)}_{\obs}$. The conditional distribution of $\bm{y}^{(i)}_{\lat}$ follows from \eqref{eq:poststates} and \eqref{eq:truncGauss} with $\bm{S}_{\lat}$ and $\bm{S}_{\obs}$ replaced by $\bm{S}^{(i)}_{\lat}$ and $\bm{S}^{(i)}_{\obs}$. Its precision matrix has dimension at most $T+P$ and bandwidth $P$, so that the operations on it are linear in $T$, and one sweep of the sampler cycles through the blocks $i = 1, \ldots, n$. Blocks of several variables can be formed in the same way.

\paragraph{Sampling from the truncated normal distribution.} Draws from \eqref{eq:truncGauss}, jointly or by blocks, can be generated by Gibbs sampling one element at a time, by the minimax exponential tilting (MET) method of \citet{botev2017normal}, or by Hamiltonian Monte Carlo (HMC). Gibbs sampling mixes slowly when the elements are strongly correlated, as consecutive values of a latent variable are here. MET produces independent draws by accept-reject sampling, but its acceptance rate falls with the dimension of the constrained vector, which here is of order $T$ for a single variable, and it is infeasible for the larger designs considered in later sections. We therefore use HMC.

HMC borrows an idea from physics \citep{Bishop06, Neal11}. It treats the current draw as the position of a particle and the negative log density as its potential energy, a landscape whose valleys are the regions of high probability. Each step gives the particle a random momentum, and hence kinetic energy, and moves it under Hamiltonian dynamics, which conserve the total energy. The particle rolls toward the valleys, exchanging potential for kinetic energy on the way, and its position after a fixed travel time is the new draw. This draw can be far from the previous one, so that consecutive draws are much less dependent than in a Gibbs sampler. Because the total energy is fixed, the particle can climb only as far as its kinetic energy allows, so it stays in the regions of high probability while it travels.

\citet{pakman2014exact} show that for a truncated normal target the Hamiltonian dynamics can be solved exactly: the potential energy is the quadratic form $\frac{1}{2}(\bm{y}_{\lat}-\bm{\mu}_{\lat})'\bm{V}_{\lat}^{-1}(\bm{y}_{\lat}-\bm{\mu}_{\lat})$, the truncation bounds act as reflecting walls, and a trajectory is followed for a fixed travel time and reflected whenever a coordinate reaches its bound. We use the Zigzag variant of \citet{nishimura2020discontinuous} and \citet{nishimura2025zigzag}, in which the momenta have Laplace instead of Gaussian distributions, so that the trajectories are piecewise linear and the reflection times are available in closed form; each step then requires only products of the banded precision matrix with a vector, and, because the dynamics are solved exactly, the accept-reject step that corrects the discretization error of HMC in general is unnecessary. \citet{zhang2022hdtg} compare these samplers, and we use their implementation.

The only tuning parameter is the travel time, which we set during the burn-in period from the smallest eigenvalue of the precision matrix of the block and keep fixed afterwards.\footnote{The travel time is seeded at $\sqrt{2}/\sqrt{\lambda_{\min}}$, where $\lambda_{\min}$ is the smallest eigenvalue, computed by the Lanczos method. During the burn-in period we refresh the eigenvalue estimate every ten sweeps with two steps of a warm-started inverse power iteration and the associated Rayleigh quotient, clamp the implied travel time to the interval $[0.01, 10]$, and use the average of the last five travel times.} HMC trades some autocorrelation of the draws for a large reduction in computing time. In Section~\ref{s:simevidence} we compare the computational cost of the samplers and report the effective sample sizes of the latent states.

\subsection{Priors}\label{sec:priors}
We assume a global-local shrinkage prior on the VAR lag coefficients. Collecting these coefficients in $\bm{\beta}$, we specify the horseshoe prior \citep{carvalho2010horseshoe}:
\begin{equation*}
    \beta_j \mid \tau,\lambda_j \sim \mathcal{N}(0,\tau^2\lambda_j^2), \qquad
    \tau \sim \mathcal{C}^{+}(0,1), \qquad
    \lambda_j \sim \mathcal{C}^{+}(0,1), \qquad j=1,\ldots,n^2P.
\end{equation*}
The common global scale shrinks coefficients towards zero, while coefficient-specific local scales allow important coefficients to escape strong shrinkage. The intercepts are excluded from hierarchical shrinkage and assigned independent $\mathcal{N}(0,1)$ priors.

For the stochastic volatility specification, we assign independent standard normal priors to the elements of $\bm{A}_0$. The unconditional means of the log-volatilities are fixed at zero, leaving the overall scale to $\bm{A}_0$. For the persistence and innovation variance parameters, we specify $(\phi_i+1)/2 \sim \mathcal{B}(5,1.5)$ and $\omega_i^2 \sim \mathcal{G}(1/2, 5)$, where the Gamma distribution uses the shape-rate parameterization; the latter is equivalent to $\omega_i \sim \mathcal{N}(0, 0.1)$. Initial log-volatilities follow their stationary distributions.

When the outlier component is included, each structural shock has a separate scale factor $o_{it}$, following \citet{carriero2021addressing}. Conditional on its outlier probability $\mathfrak{p}_i$, this factor equals one with probability $1-\mathfrak{p}_i$ and otherwise follows a discrete uniform distribution over nine equally spaced points between 2 and 20. We specify $\mathfrak{p}_i\sim\mathcal{B}(2.5,2.5(T-1))$, implying approximately one outlier per structural shock over the estimation sample a priori. The outlier factors are estimated, so the data decide which months receive a larger shock variance and by how much. A large factor reduces the influence of that month on the VAR coefficients and on the persistent volatility path, while the observations of the month remain in the likelihood and continue to enter the conditional means of later months through the lags.

\section{Simulation Evidence}\label{s:simevidence}

In this section we first describe the Monte Carlo designs, which are common to the two exercises that follow. We then compare the computational cost of the posterior samplers for the latent variables across the designs. Finally, we compare the proposed model with a Gaussian Bayesian VAR (BVAR) fitted directly to the observed data, in terms of the accuracy of one-step-ahead forecasts and of the distortion of the parameter estimates that results from treating the restricted variables as continuous.

\subsection{Simulation Design}\label{sec:design}
We generate data from a latent VAR(4) with stochastic volatility. Lag coefficients are drawn at random, with large own-lag coefficients and small cross-variable coefficients at the first lag and sparse higher lags whose scale decays with the lag, and are rescaled so that the companion matrix has a spectral radius drawn uniformly from $[0.85,0.98]$. Innovations have covariance $\bm{A}_0^{-1}\bm{H}_t\bm{A}_0^{-1}{}'$, where $\bm{A}_0$ is generated at random and the log-volatilities $h_{it}$ follow independent Gaussian AR(1) processes with mean zero, persistence $0.95$ and innovation variance $0.05$. The system is then standardized so that, at the mean log-volatility, each latent variable has unit unconditional variance. The intercepts are set so that the unconditional means of the latent variables equal $\Phi^{-1}(0.3)$, $1$, $\Phi^{-1}(0.7)$ and zero for the binary, count, censored and continuous variables, respectively. The binary, count and censored observations are obtained from the latent variables through the measurement equations \eqref{eq:linkbinary} to \eqref{eq:linkcount}, with $\psi_i = 1$ for the count variables, which are therefore generated and estimated with unit intervals; the continuous variables are observed directly.

We consider sample sizes $T\in\{100,250,750\}$, system dimensions $n\in\{10,20,30\}$, and $r\in\{1,2,3\}$ variables of each of the binary, count and censored types, retaining designs with at least two unrestricted continuous variables, which excludes $(n, r) = (10, 3)$ and leaves 24 designs. After discarding 500 initial observations, we retain $T$ observations for estimation and one for forecast evaluation. Datasets are redrawn until each binary series has a proportion of ones strictly between $0.1$ and $0.5$, each count series contains at least five distinct values, and each censored series has more than $10\%$ zeros and more than $50\%$ positive observations. These checks use all $T+1$ retained observations, including the holdout; the results therefore describe the simulation population selected by these restrictions.

The designs are used in two exercises. The timing exercise in Section~\ref{sec:comp} times single runs of the proposed model, with 1,000 burn-in and 1,000 retained draws, on one dataset per design and volatility specification; the data for the homoskedastic specification are generated with constant log-volatilities. The Monte Carlo study in Section~\ref{sec:sim} uses 500 replications per design. On each dataset it estimates the proposed model, with stochastic volatility but without the outlier component and with the equation-by-equation HMC update, and a Gaussian BVAR with the same volatility specification fitted directly to the observed outcomes, with 9,000 burn-in draws and 2,000 retained draws after thinning by three, and it compares the two models on the replications that both completed. Both models have $P = 4$ lags.

\subsection{Computational Cost}\label{sec:comp}
Table~\ref{tab:comp-time} reports the computation time per 100 iterations of the full MCMC sweep, including initialization, for each design and for four variants of the latent-state update: minimax exponential tilting (MET) and Zigzag HMC, each with joint and equation-by-equation updates, under the stochastic volatility specification and the homoskedastic specification defined in Appendix~\ref{app:emp}. Each run has a budget of one hour for its 2,000 iterations, and blank cells mark runs that did not complete within the budget.

\begin{table}[!htbp]
\centering
\footnotesize
\caption{Computation time per 100 MCMC iterations, in seconds, for the MET and Zigzag HMC samplers with joint and equation-by-equation updates across the simulation designs.}
\label{tab:comp-time}
\setlength{\tabcolsep}{2pt}
\begin{threeparttable}
\begin{tabular*}{\linewidth}{@{\extracolsep{\fill}}rrr*{8}{r}@{}}
\toprule
 & & & \multicolumn{4}{c}{Homoskedastic} & \multicolumn{4}{c}{Stochastic volatility} \\
\cmidrule(lr){4-7}\cmidrule(lr){8-11}
 &  &  & \multicolumn{2}{c}{MET} & \multicolumn{2}{c}{HMC} & \multicolumn{2}{c}{MET} & \multicolumn{2}{c}{HMC} \\
\cmidrule(lr){4-5}\cmidrule(lr){6-7}\cmidrule(lr){8-9}\cmidrule(lr){10-11}
$T$ & $n$ & $r$ & joint & eq. & joint & eq. & joint & eq. & joint & eq. \\
\midrule
100 & 10 & 1 & 14.8 & 4.28 & \textbf{1.01} & 1.16 & 13.0 & 5.41 & \textbf{1.65} & 1.86 \\
 &  & 2 &  & 7.92 & 1.78 & \textbf{1.69} &  & 8.94 & 3.00 & \textbf{2.48} \\
\addlinespace[2pt]
 & 20 & 1 & 9.43 & 5.13 & \textbf{2.38} & 2.66 & 9.36 & 6.33 & \textbf{3.76} & 4.11 \\
 &  & 2 & 108 & 8.34 & \textbf{3.16} & 3.50 &  & 11.9 & \textbf{4.49} & 4.79 \\
 &  & 3 &  & 11.4 & 4.18 & \textbf{4.15} &  & 15.5 & 6.72 & \textbf{5.72} \\
\addlinespace[2pt]
 & 30 & 1 & 10.8 & 8.18 & \textbf{5.24} & 5.78 & 14.0 & 10.6 & \textbf{7.40} & 7.97 \\
 &  & 2 & 39.3 & 11.6 & \textbf{6.27} & 6.90 & 42.6 & 13.9 & \textbf{8.50} & 9.27 \\
 &  & 3 &  & 15.2 & \textbf{7.79} & 7.95 &  & 17.3 & \textbf{10.1} & 10.5 \\
\midrule
250 & 10 & 1 &  & 109 & 2.46 & \textbf{1.82} &  & 26.7 & 3.62 & \textbf{2.95} \\
 &  & 2 &  & 66.6 & 8.98 & \textbf{2.87} &  &  & 23.8 & \textbf{4.37} \\
\addlinespace[2pt]
 & 20 & 1 & 89.7 & 20.3 & 3.91 & \textbf{3.87} & 67.8 & 19.1 & 5.77 & \textbf{5.74} \\
 &  & 2 &  &  & 8.39 & \textbf{5.70} &  &  & 10.1 & \textbf{7.66} \\
 &  & 3 &  &  & 20.9 & \textbf{7.57} &  & 161 & 29.5 & \textbf{9.31} \\
\addlinespace[2pt]
 & 30 & 1 & 107 & 25.7 & 7.96 & \textbf{7.88} & 106 & 31.0 & \textbf{10.7} & 10.8 \\
 &  & 2 &  & 38.8 & 12.5 & \textbf{10.7} &  & 42.4 & 15.1 & \textbf{13.2} \\
 &  & 3 &  & 52.6 & 19.7 & \textbf{13.4} &  & 59.4 & 24.3 & \textbf{16.4} \\
\midrule
750 & 10 & 1 &  &  & 14.7 & \textbf{5.98} &  &  & 17.6 & \textbf{8.30} \\
 &  & 2 &  &  & 95.8 & \textbf{11.2} &  &  & 154 & \textbf{15.3} \\
\addlinespace[2pt]
 & 20 & 1 &  &  & 15.1 & \textbf{9.75} &  &  & 18.0 & \textbf{12.9} \\
 &  & 2 &  &  & 53.6 & \textbf{17.5} &  &  & 62.3 & \textbf{22.2} \\
 &  & 3 &  &  &  & \textbf{25.1} &  &  & 164 & \textbf{27.8} \\
\addlinespace[2pt]
 & 30 & 1 &  &  & 22.2 & \textbf{17.7} &  &  & 26.0 & \textbf{22.4} \\
 &  & 2 &  &  & 53.7 & \textbf{27.0} &  &  & 62.3 & \textbf{32.6} \\
 &  & 3 &  &  & 132 & \textbf{35.8} &  &  & 135 & \textbf{42.4} \\
\bottomrule
\end{tabular*}
\begin{tablenotes}[flushleft]\footnotesize
\item \textit{Notes:} $T$: observations; $n$: variables; $r$: variables of each binary, count, and censored type. Each design includes at least two continuous variables. MET: minimax exponential tilting; HMC: Hamiltonian Monte Carlo; joint and eq.: joint and equation-by-equation updates. Bold marks the fastest method for each design and volatility specification. Entries are seconds per 100 iterations: the total time of each run, including initialization, divided by 20, since each run has 2,000 iterations (1,000 burn-in and 1,000 retained). Blank cells indicate that the run did not complete within one hour, so no comparable average is reported.
\end{tablenotes}
\end{threeparttable}
\end{table}

We make three observations. First, HMC is the fastest method in every design, and the equation-by-equation HMC update completes every design; in the largest design, with $T = 750$, $n = 30$ and nine restricted variables, it requires about 42 seconds per 100 iterations under stochastic volatility. Second, MET does not scale: the joint MET update fails to complete within the budget in most designs with $T = 250$, and no MET variant completes any design with $T = 750$. Third, the joint HMC update is marginally faster in most of the small designs with $T = 100$, whereas the equation-by-equation update is faster in nearly every design with $T = 250$ and in every design with $T = 750$, and its advantage grows with $T$ and with the number of restricted variables. Each of its blocks has dimension at most $T+P$ and bandwidth $P$, whereas both the dimension and the bandwidth of the joint update grow with the number of restricted variables. 

We note that the reported times are not times per effective draw: MET produces independent draws from the conditional distribution of the latent states, whereas the HMC draws are autocorrelated, with inefficiency factors between 1.4 and 6.1 in the Monte Carlo study below (Table~\ref{tab:sim}). In the designs with $T = 100$, where MET completes, the comparison per effective draw is closer and in some designs favors MET; in the larger designs MET does not complete within the budget, so HMC is the only feasible option.

\subsection{Estimation and Forecast Accuracy}\label{sec:sim}

Binary observations identify the sign of the latent variable but leave its scale unrestricted. For evaluation, we apply the normalization of Section~\ref{sec:norm} to the data generating process (DGP) and to the posterior draws of both models, so that each binary latent equation has unit time-average innovation variance $v_i = [\bm{\bar{\Sigma}}]_{ii}$ over the estimation sample, excluding the initial four lags and the holdout. With $\bm{D} = \diag(d_1,\hdots,d_n)$, where $d_i = \sqrt{v_i}$ for the binary variables and $d_i = 1$ otherwise, the lag matrices, the intercepts and the covariance matrices are transformed as $\bm{D}^{-1}\bm{A}_p\bm{D}$, $\bm{D}^{-1}\bm{a}$ and $\bm{D}^{-1}\bm{\Sigma}_t\bm{D}^{-1}$, and the latent forecasts are premultiplied by $\bm{D}^{-1}$. This is the full transformation implied by dividing the binary latent variables by $\sqrt{v_i}$. The draws of the proposed model satisfy the normalization by construction, so that $\bm{D} = \bm{I}_n$ and they are unchanged; the transformation therefore matters for the DGP and for the draws of the BVAR, and it ensures that the losses of both models are free of the arbitrary scale of the binary latent variables. This convention fixes the average innovation variances; the innovation variances at individual dates and the unconditional variances of the latent levels remain free. The fixed count thresholds and the observed continuous values anchor the remaining scales. Since the Gaussian BVAR is fitted directly to the observed outcomes, it is misspecified, and its parameter losses measure how well it recovers the parameters of the latent data generating process.

Within each replication, parameter losses are the mean absolute deviations of the elementwise posterior medians from the normalized truth, computed separately over all lag coefficients, intercepts, and unique elements of the time-average innovation covariance matrix. One-step-ahead forecasts are evaluated using the continuous ranked probability score \citep[CRPS;][]{gneiting2007strictly}, with the Brier score for the observed binary outcomes. For the BVAR, the predictive draws are mapped to the support of each outcome: they are clipped to $[0,1]$ for the binary variables, rounded upward and bounded below by zero for the count variables, and censored at zero for the censored variables. Each forecast score is divided by the corresponding variable's training-sample standard deviation on the latent or observed scale, then averaged across variables within the reported group.

\begin{table}[!htbp]
\centering
\footnotesize
\caption{Median simulation losses relative to the Gaussian BVAR and sampling diagnostics.}
\label{tab:sim}
\setlength{\tabcolsep}{2pt}
\begin{threeparttable}
\begin{tabular*}{\linewidth}{@{\extracolsep{\fill}}rl*{8}{r}@{}}
\toprule
 &  & \multicolumn{2}{c}{$n=10$} & \multicolumn{3}{c}{$n=20$} & \multicolumn{3}{c}{$n=30$} \\
\cmidrule(lr){3-4}\cmidrule(lr){5-7}\cmidrule(lr){8-10}
$T$ &  & $r=1$ & $r=2$ & $r=1$ & $r=2$ & $r=3$ & $r=1$ & $r=2$ & $r=3$ \\
\midrule
\multicolumn{10}{@{}l}{\textit{A. Parameter estimation: relative MAE}} \\
\addlinespace[2pt]
100 & Dynamics & 1.007 & 1.000 & 0.992 & 0.991 & 0.991 & 0.990 & 0.986 & 0.978 \\
 & Intercepts & 0.546 & 0.492 & 0.652 & 0.546 & 0.523 & 0.730 & 0.606 & 0.567 \\
 & Covariance & 0.857 & 0.782 & 0.942 & 0.911 & 0.905 & 0.980 & 0.958 & 0.950 \\
\addlinespace[3pt]
250 & Dynamics & 0.964 & 0.943 & 0.983 & 0.978 & 0.986 & 0.998 & 0.988 & 0.989 \\
 & Intercepts & 0.420 & 0.392 & 0.557 & 0.480 & 0.469 & 0.619 & 0.539 & 0.515 \\
 & Covariance & 0.688 & 0.525 & 0.822 & 0.728 & 0.675 & 0.895 & 0.820 & 0.763 \\
\addlinespace[3pt]
750 & Dynamics & 0.894 & 0.849 & 0.955 & 0.930 & 0.924 & 0.980 & 0.962 & 0.958 \\
 & Intercepts & 0.306 & 0.287 & 0.432 & 0.398 & 0.388 & 0.526 & 0.461 & 0.456 \\
 & Covariance & 0.493 & 0.331 & 0.645 & 0.485 & 0.397 & 0.740 & 0.595 & 0.504 \\
\midrule
\multicolumn{10}{@{}l}{\textit{B. Forecast evaluation: relative scores}} \\
\addlinespace[2pt]
100 & Latent: all & 0.855 & 0.762 & 0.936 & 0.888 & 0.846 & 0.960 & 0.932 & 0.904 \\
 & Binary & 0.845 & 0.915 & 0.625 & 0.860 & 0.892 & 0.628 & 0.852 & 0.906 \\
 & Count & 0.813 & 0.793 & 0.867 & 0.782 & 0.822 & 0.820 & 0.768 & 0.810 \\
 & Censored & 0.923 & 0.921 & 0.968 & 0.987 & 0.956 & 0.967 & 0.931 & 0.954 \\
 & Continuous & 1.007 & 0.986 & 1.001 & 0.994 & 0.990 & 0.999 & 1.001 & 0.991 \\
 & Observed: all & 0.958 & 0.907 & 0.986 & 0.977 & 0.964 & 0.989 & 0.984 & 0.971 \\
\addlinespace[3pt]
250 & Latent: all & 0.827 & 0.730 & 0.915 & 0.865 & 0.837 & 0.940 & 0.909 & 0.871 \\
 & Binary & 1.070 & 1.093 & 0.681 & 0.906 & 1.010 & 0.726 & 0.831 & 0.907 \\
 & Count & 0.714 & 0.793 & 0.816 & 0.873 & 0.870 & 0.790 & 0.788 & 0.798 \\
 & Censored & 0.946 & 0.973 & 0.981 & 0.927 & 0.997 & 0.989 & 0.926 & 0.972 \\
 & Continuous & 1.011 & 0.991 & 0.995 & 0.994 & 1.000 & 0.995 & 0.991 & 1.005 \\
 & Observed: all & 0.954 & 0.913 & 0.978 & 0.939 & 0.959 & 0.979 & 0.975 & 0.969 \\
\addlinespace[3pt]
750 & Latent: all & 0.833 & 0.649 & 0.890 & 0.820 & 0.765 & 0.939 & 0.868 & 0.818 \\
 & Binary & 0.830 & 1.100 & 0.785 & 0.862 & 0.949 & 0.727 & 0.820 & 0.903 \\
 & Count & 0.917 & 0.729 & 0.790 & 0.730 & 0.746 & 0.730 & 0.765 & 0.733 \\
 & Censored & 0.970 & 0.937 & 1.050 & 0.933 & 0.954 & 0.998 & 0.943 & 0.959 \\
 & Continuous & 1.000 & 0.966 & 1.005 & 1.010 & 0.989 & 1.002 & 0.994 & 0.987 \\
 & Observed: all & 0.955 & 0.899 & 0.980 & 0.935 & 0.926 & 0.989 & 0.976 & 0.936 \\
\midrule
\multicolumn{10}{@{}l}{\textit{ProToC-VAR sampling diagnostics}} \\
\addlinespace[2pt]
100 & ESS (IF) & 1052 (1.9) & 1019 (2.0) & 588 (3.4) & 582 (3.4) & 558 (3.6) & 338 (5.9) & 337 (5.9) & 327 (6.1) \\
250 & ESS (IF) & 1173 (1.7) & 1095 (1.8) & 775 (2.6) & 730 (2.7) & 727 (2.8) & 505 (4.0) & 485 (4.1) & 499 (4.0) \\
750 & ESS (IF) & 1412 (1.4) & 1258 (1.6) & 1164 (1.7) & 973 (2.1) & 901 (2.2) & 872 (2.3) & 733 (2.7) & 668 (3.0) \\
\bottomrule
\end{tabular*}
\begin{tablenotes}[flushleft]\footnotesize
\item \textit{Notes:} $T$: observations; $n$: variables; $r$: variables of each binary, count, and censored type. Each design includes at least two continuous variables. Entries in panels A and B are ratios of median losses across 500 replications, ProToC-VAR relative to the Gaussian BVAR fitted to the observed outcomes. In panel A and in the row Latent: all, the losses are measured against the latent data generating process, so the ratios quantify the distortion from treating the restricted variables as continuous; in the remaining rows of panel B, both models forecast the same observed outcomes and values below one favor ProToC-VAR. MAE: mean absolute error. Forecast rows refer to observed outcomes except for Latent: all. Forecast scores are divided by training-sample standard deviations: Brier scores for binary outcomes, continuous ranked probability scores otherwise. For each ProToC-VAR replication, ESS is the median latent-state effective sample size and IF (inefficiency factor) is $2000/\mathrm{ESS}$, based on 2000 saved draws. The table reports medians of both diagnostics across replications.
\end{tablenotes}
\end{threeparttable}
\end{table}

Table~\ref{tab:sim} reports ratios of median losses, ProToC-VAR relative to the BVAR, taking medians separately for each model and metric across replications. In panel A, and in the latent-scale forecast row of panel B, the losses of both models are measured against the latent data generating process, so the ratios quantify the distortion that results from treating the restricted variables as continuous; in the remaining rows of panel B, both models forecast the same observed outcomes, and ratios below one favor ProToC-VAR. The table also reports the median latent-state effective sample size (ESS) and inefficiency factor (IF) across ProToC-VAR replications. Within each replication, ESS is summarized by its median across latent states.

We note three patterns in Table~\ref{tab:sim}. First, treating the restricted variables as continuous distorts the intercepts and the innovation covariance matrix in every design: the loss ratios lie between 0.29 and 0.73 for the intercepts and between 0.33 and 0.98 for the covariance matrix, and they fall as $T$ grows, because the estimation error of the proposed model shrinks while the distortion of the BVAR does not. Second, the lag coefficients are much less affected: their loss ratios are close to one at $T = 100$ and lie between 0.85 and 0.98 at $T = 750$. Third, in forecasting the observed outcomes, the proposed model improves on the BVAR for the count variables in every design and for the binary and censored variables in most designs, while the two models are on par for the continuous variables. Averaged over all observed variables (the row Observed: all), the proposed model is ahead in every design. Overall, these results show that the shortcut of treating restricted variables as continuous distorts the levels and the innovation scales of the latent system, and that modeling the observation mechanism improves the forecasts of the restricted outcomes at no cost for the continuous variables.

\section{Modeling US Macro-Financial Dynamics}\label{s:application}

We now apply the proposed model to 25 monthly US macro-financial variables, among them the NBER recession indicator, the number of bank failures and six interest rates that are subject to censoring at the effective lower bound. After describing the data, we examine the estimates of the latent variables underlying the restricted series, the imputed early observations of two variables and the time-varying volatilities. We then use the model for conditional forecasting and scenario analysis with conditioning assumptions on restricted and continuous variables alike.

\subsection{Data}
Our monthly dataset runs from 1960:01 to 2025:12 and is sourced from FRED-MD \citep{mccracken2016fred} for most macroeconomic and financial variables. The binary variable \texttt{NBERREC} is the indicator of expansions and recessions (contractions) released by the \href{https://www.nber.org/research/data/us-business-cycle-expansions-and-contractions}{NBER Business Cycle Dating Committee}. The count variable \texttt{BANKFAIL} is the number of bank failures, aggregated from the individual bank-level data on \href{https://banks.data.fdic.gov/bankfind-suite/failures}{banks.data.fdic.gov}. The censored variables are six interest rates at maturities from overnight to ten years, some of which have been at their ELB during the sample. Following \citet{carriero2023shadow}, we treat all observations at or below $0.25$ percent as constrained by the ELB and record them at the threshold, so that \texttt{FEDFUNDS}, \texttt{TB3MS}, \texttt{TB6MS} and \texttt{GS1} each have at least one censored observation.

Our information set includes the key variables assessed by the NBER Business Cycle Dating Committee for dating recessions---non-farm payroll employment (\texttt{PAYEMS}), industrial production (\texttt{INDPRO}), real manufacturing and trade sales (\texttt{CMRMTSPLx}), real personal income excluding transfers (\texttt{W875RX1})---and adds the federal funds rate (\texttt{FEDFUNDS}) and a term spread (\texttt{TS10G1}) as important financial predictors. This information set also mirrors and extends the one of \citet{mccracken2022binary}.

The remaining unrestricted variables cover the labor market (the unemployment rate, and average weekly hours and hourly earnings in goods-producing industries), prices (the PCE and CPI price indices and the crude oil price), housing starts, and financial markets (the S\&P 500 index, the US--UK exchange rate, Moody's Baa corporate bond yield, the excess bond premium of \citet{gilchrist2012credit} and the Chicago Fed national financial conditions index). The last two are available from 1973:01 and 1971:01 onwards, and their earlier values are treated as missing.

Employment, industrial production, sales, real income, hourly earnings, the price indices, the stock price index, the exchange rate and the oil price enter as monthly log-differences in percent; weekly hours and housing starts enter in logs; and the remaining variables enter in levels. The transformations follow \citet{mccracken2022binary} for the real activity and price variables and \citet{carriero2023shadow} for the interest rates, so that the information set is comparable with theirs. One could instead use the variables in levels, with the same shrinkage prior, since the measurement equations do not depend on how the continuous variables are transformed. We keep the growth-rate specification because the objects reported below, the latent states and the conditional forecasts over a three-year horizon, depend on the short-run dynamics of the system. Appendix~\ref{app:emp} lists all variables with their types and transformations.

\subsection{Estimates of the Latent States} \label{sec:latent_states}

We estimate the model with $P = 12$ lags, stochastic volatility and the outlier component in all equations. The count variable uses the log thresholds ($\psi_i = 0$), the interest rates are censored at $0.25$ percent, and we apply the normalizations of Section~\ref{sec:norm}. Before estimation, the continuous and censored series are standardized using outlier-robust measures of location and scale, and the count series and its thresholds are divided by the standard deviation of $\log(k + 1/2)$ over the months with positive counts $k$, which leaves the zero threshold at zero; all estimates are transformed back to the original units. The latent states are sampled with the equation-by-equation Zigzag HMC update. We run the sampler for 65,000 iterations, discard the first 15,000 as burn-in, and retain every 10th of the remaining draws, which gives 5,000 posterior draws.

A key output of the proposed VAR is the set of latent states associated with the restricted variables. The state underlying the NBER recession indicator can be interpreted as an indicator of the business cycle and its turning points, the state underlying the bank failure count as the intensity of banking distress, and the states associated with the interest rates, or shadow rates, as a measure of the overall stance of monetary policy when the ELB is binding. The model also imputes the missing early observations of the two unrestricted variables that start after 1960, and it delivers the time-varying volatilities and the outlier component of all innovations.

\begin{figure}[!htbp]
    \centering
    \includegraphics[width=\linewidth]{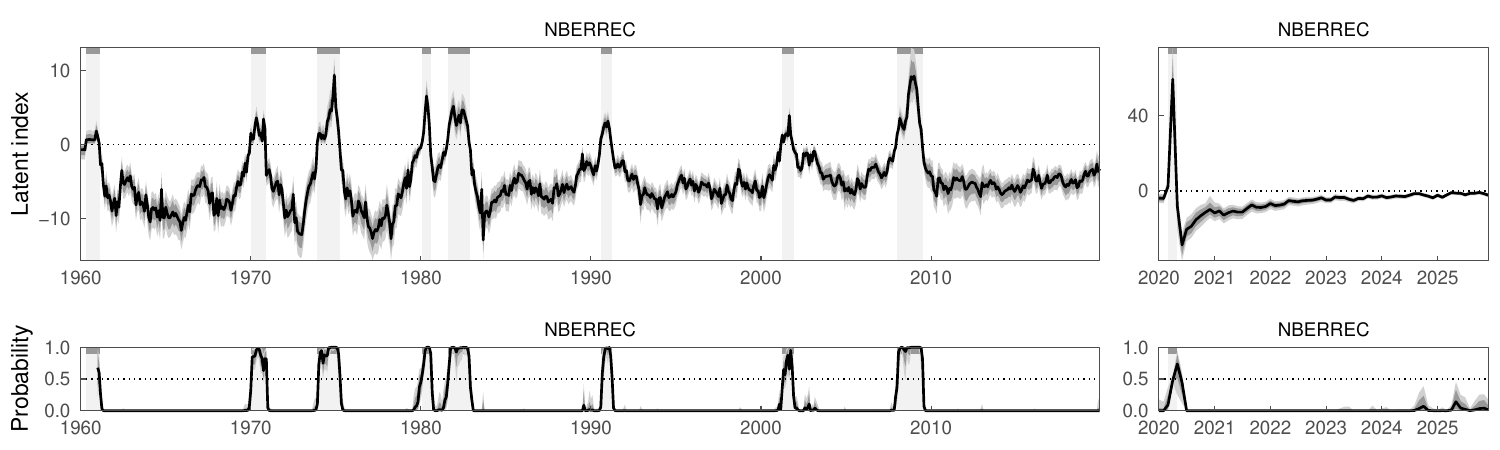}
    \caption{Estimates of the business cycle indicator underlying \texttt{NBERREC} (upper panels) and the implied recession probabilities (lower panels).}
    \label{fig:latent_binary}
    \fnote{\textit{Notes:} Posterior median alongside 68/90 percent credible sets. The indicator is counter-cyclical and rises above zero in recessions. Shaded areas indicate NBER recessions. The left panels cover 1960:01 to 2019:12, the right panels the period from 2020:01 onwards; the samples are plotted separately so that the COVID-19 observations do not compress the scale of the business cycle indicator for the remainder of the sample.}
\end{figure}

Figure~\ref{fig:latent_binary} presents the dynamic evolution of the business cycle indicator and the corresponding recession probabilities. The recession probability at date $t$ is $\Phi(\mu_{b,t}/\sigma_{b,t})$, where $\mu_{b,t}$ is the conditional mean of the business cycle indicator given the lagged variables and $\sigma_{b,t}$ its innovation standard deviation at date $t$, including the outlier component; the figure shows posterior quantiles of this probability across the draws. Our implementation gives a counter-cyclical indicator, which is negative throughout expansions and rises above zero in every NBER recession; a pro-cyclical indicator can be obtained by redefining the binary variable. As discussed in \citet{dueker2010forecasting}, the distance of the indicator from zero measures how far the business cycle is from its next turning point. By this metric, the most severe recessions in the sample before 2020 are the 1973--75 recession and the Great Recession of 2007--09. The implied recession probabilities switch between values close to zero and one within a few months of the NBER turning points, and the credible sets are narrow relative to the movements of the indicator. The COVID-19 recession produces a spike in the indicator about six times as high as the earlier peaks, which the outlier component absorbs (see Figure~\ref{fig:latent_sigma} below); the right panels therefore show this period on its own scale. After the 2020 recession the indicator falls to the lowest value of the sample, which reflects the record growth rates of employment, industrial production, sales and income during the recovery, and it then rises gradually, with brief and uncertain increases in the recession probability in 2024 and 2025.

\begin{figure}[!htbp]
    \centering
    \includegraphics[width=\linewidth]{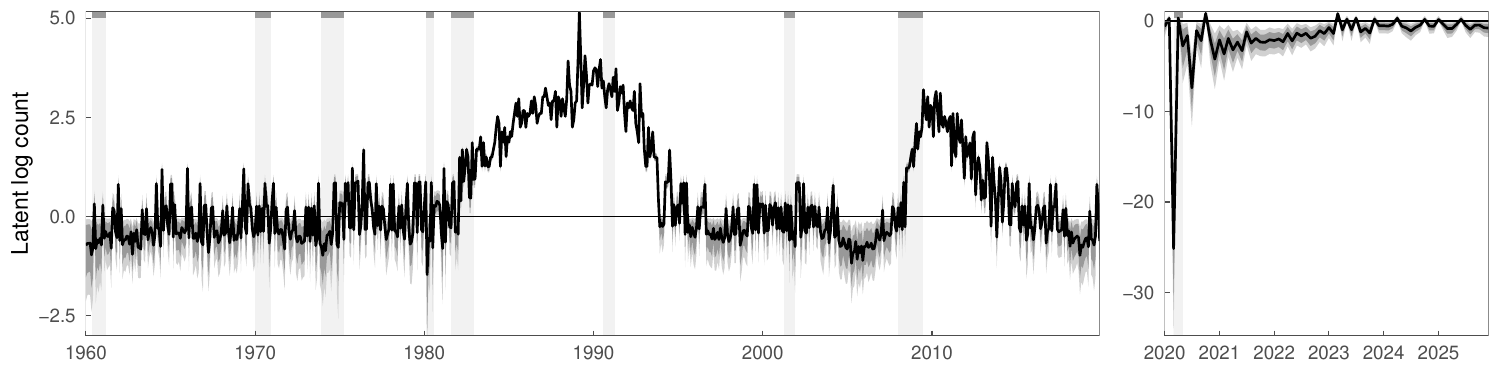}
    \caption{Estimates of the intensity of banking distress underlying the number of FDIC-insured bank failures (\texttt{BANKFAIL}).}
    \label{fig:latent_count}
    \fnote{\textit{Notes:} Posterior median alongside 68/90 percent credible sets. The intensity is on the model scale, that is, on the scale of the log thresholds ($\psi_i = 0$) in the count measurement equation. Shaded areas indicate NBER recessions; the left and right panels split the sample as in Figure~\ref{fig:latent_binary}.}
\end{figure}

Figure~\ref{fig:latent_count} shows the intensity of banking distress underlying the bank failure count on the log scale, on which values below zero correspond to months without failures. Two episodes stand out: the savings and loan crisis, during which the intensity rises from 1982 to a peak in 1989 and returns to zero by 1994, and the Global Financial Crisis, with a rise from 2008 to a peak in 2009 and a return to zero by 2015. In calm periods the intensity fluctuates around zero and its lower credible band is wide, because a month without failures only bounds it from above. The COVID-19 months push the intensity far below zero, another outlier episode, after which it returns to values around zero, with brief excursions above zero at the bank failures of 2023 and 2024.

\begin{figure}[!htbp]
    \centering
    \includegraphics[width=\linewidth]{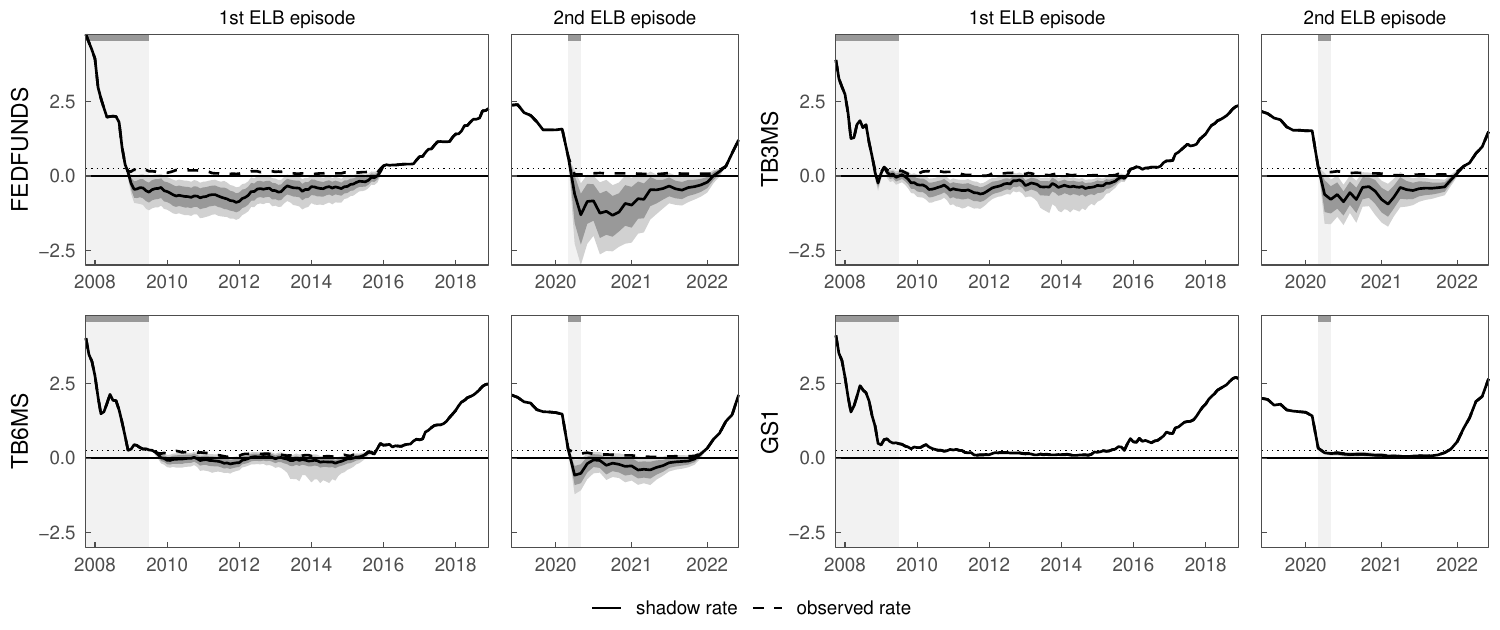}
    \caption{Estimates of the shadow rates underlying the interest rates subject to censoring at the ELB, around the two ELB episodes (2007:10 to 2018:12 and 2019:06 to 2022:06).}
    \label{fig:latent_censored}
    \fnote{\textit{Notes:} Solid lines show posterior medians of the shadow rates alongside 68/90 percent credible sets, dashed lines the observed interest rates. The panels show the four interest rates that have censored observations in the sample. The dotted horizontal line marks the $0.25$ percent censoring threshold. Shaded areas indicate NBER recessions.}
\end{figure}

Figure~\ref{fig:latent_censored} reports the shadow rates that arise when the interest rates hit the ELB. As discussed in \citet{carriero2023shadow}, the shadow rates may be viewed as the monetary policy reactions to the variables in the VAR, and they capture the overall stance of policy when the central bank implements alternative policies, such as balance sheet policies and forward guidance, to stimulate the economy. The two episodes differ in depth and duration. In the first, which begins during the Great Recession and lasts until the liftoff in late 2015, the shadow federal funds rate declines gradually to about $-0.9$ percent in 2011--12. The shadow rates are ordered by maturity, with the one-year rate staying close to the threshold, and the five- and ten-year rates are never censored. In the second, the shadow federal funds rate falls quickly to about $-1.3$ percent in 2020--21, with wider credible sets, and the shadow rates leave the ELB in early 2022 when policy tightens in response to inflation.

\begin{figure}[!htbp]
    \centering
    \includegraphics[width=\linewidth]{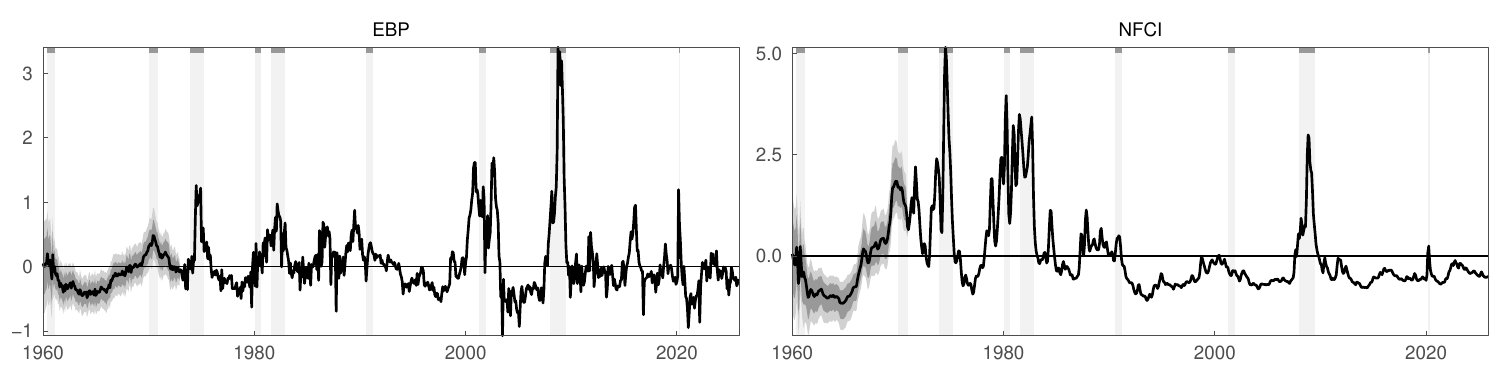}
    \caption{Imputed early observations of the excess bond premium (\texttt{EBP}) and the Chicago Fed national financial conditions index (\texttt{NFCI}), which are unavailable at the beginning of the sample.}
    \label{fig:latent_missing}
    \fnote{\textit{Notes:} Posterior median alongside 68/90 percent credible sets. \texttt{EBP} is observed from 1973:01, \texttt{NFCI} from 1971:01 onwards; earlier observations are imputed alongside the remaining latent states. Shaded areas indicate NBER recessions.}
\end{figure}

Figure~\ref{fig:latent_missing} shows the excess bond premium and the financial conditions index, which are observed from 1973:01 and 1971:01 onwards, respectively; the model treats the earlier observations as missing and imputes them alongside the other latent states. The imputed paths rise into the 1969--70 recession, consistent with the tightening of credit and financial conditions at the end of the 1960s.

Figure~\ref{fig:latent_sigma} shows the reduced-form log innovation standard deviations of the recession indicator, the bank failure count and the federal funds rate; we report the log innovation standard deviations for the other variables in Appendix~\ref{app:emp}. The innovation volatility of the federal funds rate is highest during the Volcker disinflation of 1979--82 and lowest in the 2010s, during and after the first ELB episode. The volatility of the recession indicator drifts down over the sample, and its time average is close to zero, which reflects the unit-scale normalization of Section~\ref{sec:norm}. The circles mark the months in which the outlier component is active, among them several recession months and, for all three variables, the spring of 2020. There it raises the innovation standard deviations for a few months by up to a factor of about 20 without affecting the persistent volatility component.

\begin{figure}[!htbp]
    \centering
    \includegraphics[width=\linewidth]{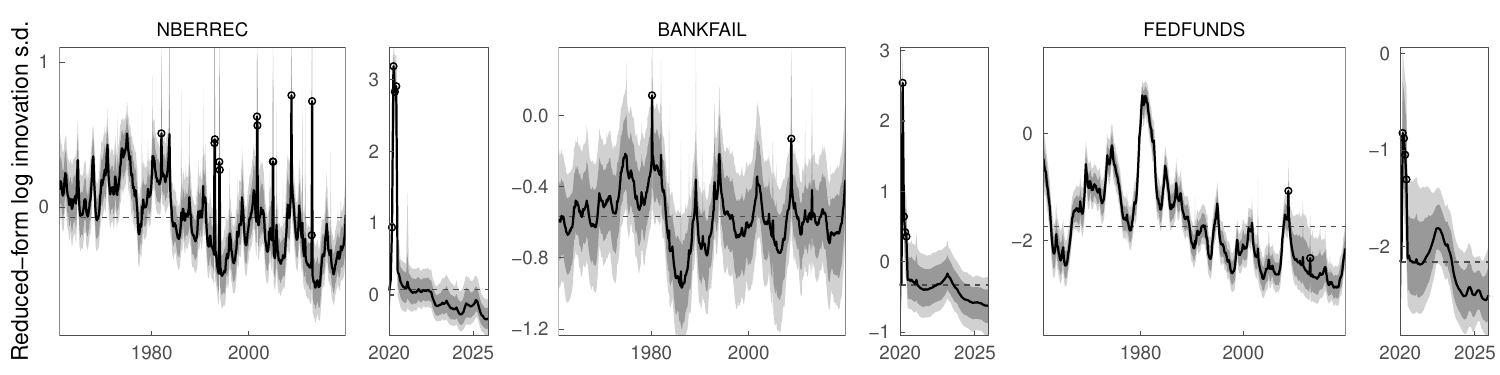}
    \caption{Estimates of the reduced-form log innovation standard deviations, $\log\sqrt{[\bm{\Sigma}_t]_{ii}}$ for \texttt{NBERREC}, \texttt{BANKFAIL} and \texttt{FEDFUNDS}.}
    \label{fig:latent_sigma}
    \fnote{\textit{Notes:} Posterior median alongside the 68/90 percent credible sets. The reduced-form variances are computed from $\bm{\Sigma}_t = \bm{A}_0^{-1}\bm{O}_t\bm{H}_t\bm{O}_t\bm{A}_0^{-1}{}'$ and thus include the outlier component; circles mark the months in which the posterior median including the outlier component exceeds the one excluding it by a meaningful amount. Dashed horizontal lines indicate the time average of the posterior median within a panel.}
\end{figure}

We repeat Figures~\ref{fig:latent_binary} to \ref{fig:latent_missing} for the homoskedastic version of the model in Appendix~\ref{app:emp}. The turning points of the business cycle indicator, the two episodes of banking distress and the imputed series are essentially unchanged, although the homoskedastic indicator separates recessions from expansions less sharply. The shadow rates differ. Without stochastic volatility, the innovation variances of the interest rates are averages over the whole sample, including the volatile years around 1980, and the shadow federal funds rate falls to about $-2.2$ percent in both ELB episodes, with wider credible sets. The low volatility that the stochastic volatility model estimates at the ELB keeps the shadow federal funds rate at about $-0.9$ and $-1.3$ percent in the two episodes, about one percentage point above its homoskedastic counterpart.

\FloatBarrier
\subsection{Conditional Forecasts and Scenario Analysis} \label{sec:cond_fc}

Conditional forecasting is a popular tool for assessing the paths of a set of variables of interest under assumptions about the future paths of other variables \citep{waggoner1999conditional, banbura2015conditional, antolin2021structural}. For example, \citet{mccracken2022binary} use it to examine the effects of monetary policy interventions and of increases in oil prices on the likelihood of a recession. We demonstrate the usefulness of the proposed framework with five conditional forecasting exercises, chosen to showcase the variable types that it accommodates: the scenarios concern the business cycle indicator, the count of bank failures and interest rates that are subject to the ELB, as well as continuous variables.

All scenarios start in 2026:01, the month after the final in-sample observation, and the forecast horizon is 36 months. The conditioning assumptions are imposed as restrictions on the joint predictive distribution derived in Appendix~\ref{app:pred}, which allows us to condition on the observed variables and on their latent counterparts; the restrictions are imposed on the means while the variances of the unconditional predictive distribution are preserved \citep[see][]{antolin2021structural}. For the binary and count variables, the restrictions are placed on the latent variables: a target probability is imposed as the corresponding quantile of the business cycle indicator, and a count path as the logarithms of the interval midpoints of the counts. These scenarios fix the expected paths of the latent variables, and the realized binary and count outcomes remain random. The outlier factors are set to one over the forecast horizon, so that the predictive distributions describe normal times and exclude the risk of further outlier episodes. For each scenario we report either the conditional forecasts of selected variables or generalized impulse response functions \citep[GIRFs; see][]{koop1996impulse}, computed draw by draw as the difference between the conditional and the unconditional forecasts, that is, the responses to the reduced-form innovations that implement the restrictions. Appendix~\ref{app:emp} reports, for the most severe variant of each scenario, the conditional forecasts of all variables, including the latent states, alongside the unconditional forecasts.

\begin{figure}[!htbp]
    \centering
    \includegraphics[width=\linewidth]{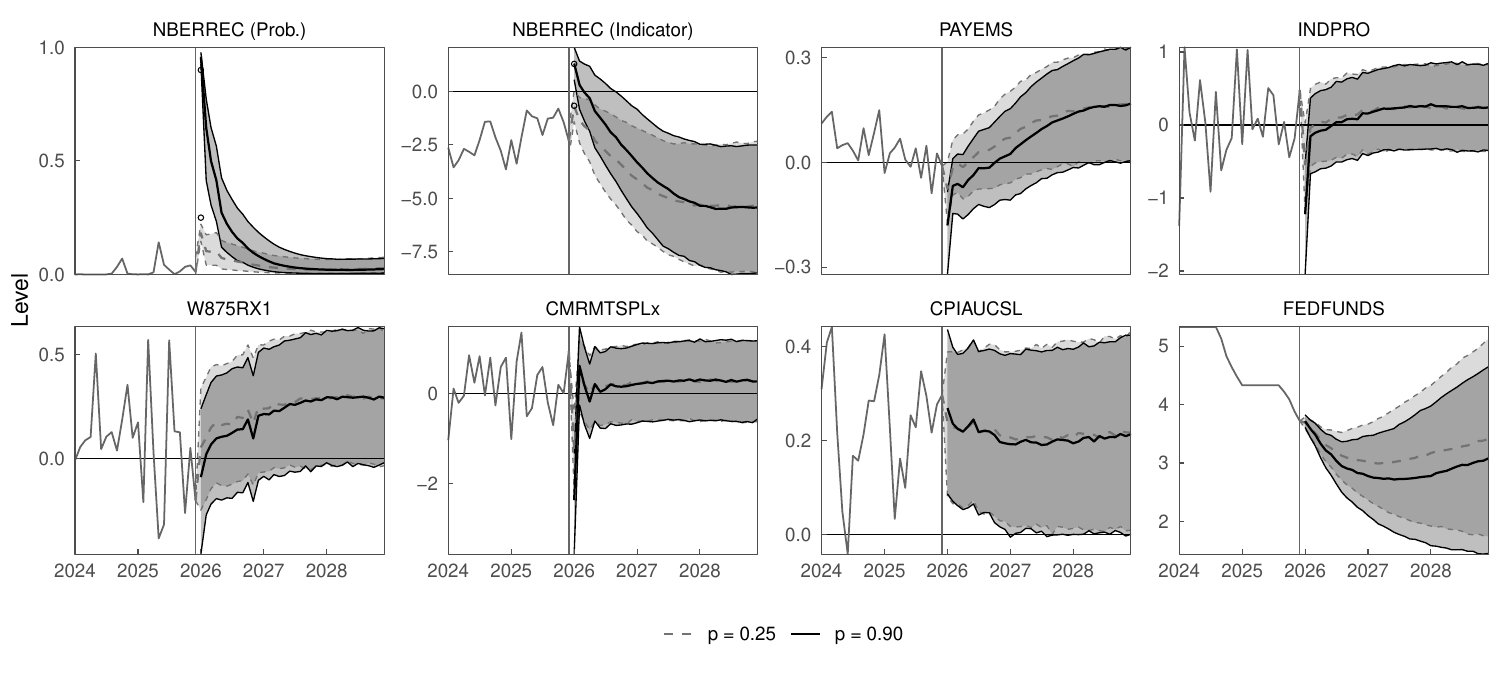}
    \caption{Recession probability scenarios: conditional forecasts when the business cycle indicator is fixed on impact at the $p = 0.25$ and $p = 0.9$ quantiles of the standard normal distribution.}
    \label{fig:cf_recession}
    \fnote{\textit{Notes:} Posterior medians alongside 68 percent credible sets of the conditional forecasts from 2026:01 to 2028:12, with the band boundaries drawn in the line type of the respective scenario; the grey line shows the in-sample posterior median for the preceding two years, up to the vertical line at the forecast origin. Circles mark the imposed restriction, shown on the probability scale for \texttt{NBERREC} (Prob.) and on the scale of the business cycle indicator for \texttt{NBERREC} (Indicator).}
\end{figure}

\textit{Recession probability scenarios.} We condition the business cycle indicator underlying \texttt{NBERREC} on impact at the quantiles $p \in \{0.25, 0.9\}$ of the standard normal distribution. Given the scale normalization of the indicator, this targets a conditional recession probability of approximately $p$ in the first month of the forecast horizon. Figure~\ref{fig:cf_recession} shows the conditional forecasts of the recession probability and of the key indicators of the NBER dating committee. Under $p = 0.9$, the recession probability falls below 0.2 within about half a year and is close to zero by 2027: in the absence of further shocks, the model expects a recession to be short, as recessions have been in the sample. Employment, industrial production, real income and sales contract on impact and recover within a year. On impact, employment growth falls by about 0.2 percent per month and industrial production by more than one percent. Inflation is essentially unaffected, and the federal funds rate falls to about 30 basis points below its path under $p = 0.25$ and stays below it over the horizon. The $p = 0.25$ scenario is close to the unconditional forecast.

\begin{figure}[!htbp]
    \centering
    \includegraphics[width=\linewidth]{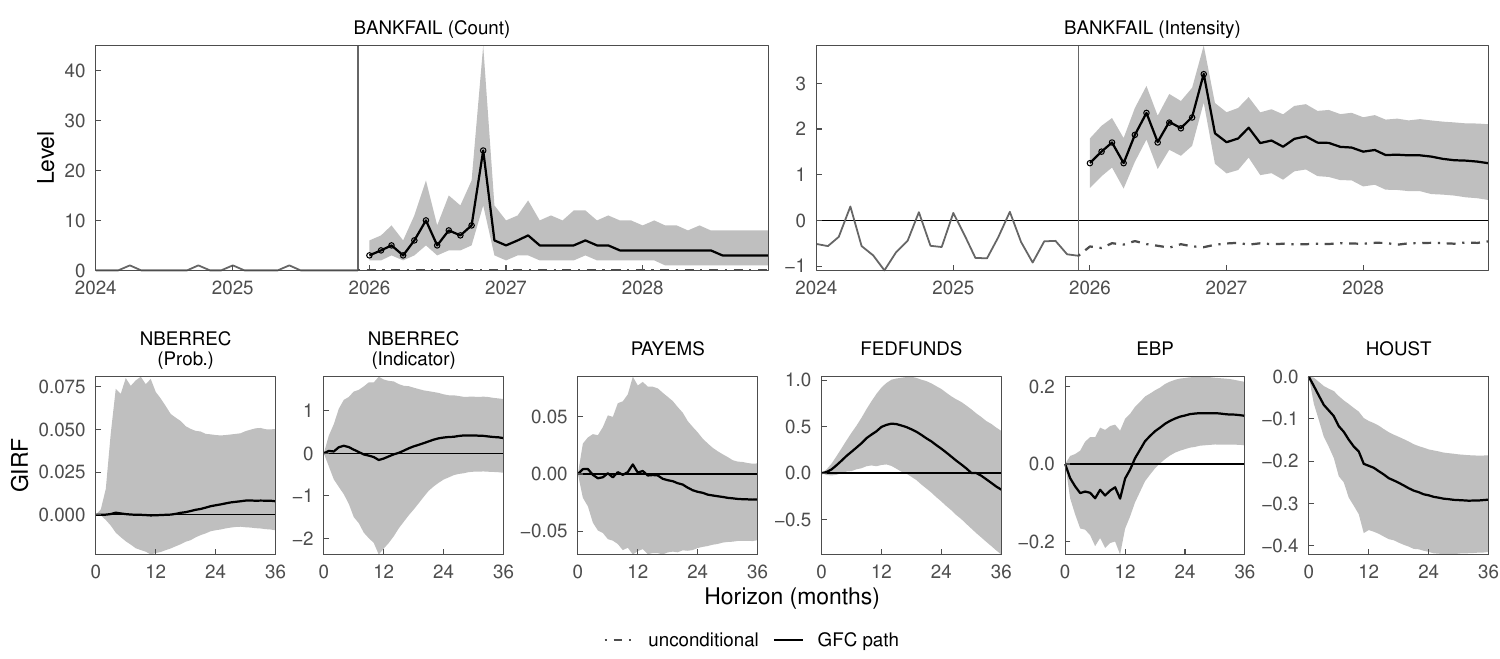}
    \caption{Bank failure scenario: the observed Global Financial Crisis path of \texttt{BANKFAIL} (2008:09 to 2009:07) is imposed from 2026:01 onwards.}
    \label{fig:cf_banks}
    \fnote{\textit{Notes:} The top row shows the imposed expected path of bank failures (circles mark the restrictions) alongside the unconditional forecast (dash-dotted), for the count and for the intensity of banking distress; posterior medians with 68 percent credible sets. The bottom row shows GIRFs, i.e., the draw-wise difference between the conditional and the unconditional predictive distribution, in units of the respective (transformed) variable against the forecast horizon.}
\end{figure}

\textit{Bank failures.} We impose the observed path of bank failures during the Global Financial Crisis, from 2008:09 to 2009:07, as the expected path of bank failures from 2026:01 onwards, which places the restriction on the latent intensity of banking distress. The top row of Figure~\ref{fig:cf_banks} shows the imposed path, which rises from a handful of failures to 24 per month, alongside the unconditional forecast and the implied intensity of banking distress; the bottom row shows the GIRFs of selected variables. Housing starts fall by about 25 percent within two years and remain depressed, employment growth declines modestly in the second and third years, the excess bond premium rises after a year, and the business cycle indicator rises by about 0.4 units, which raises the recession probability only slightly from its low unconditional level. The federal funds rate rises by about 50 basis points over the first year and falls below its unconditional path at the end of the horizon. The GIRFs are responses to reduced-form innovations, and the response of the policy rate reflects the historical pattern in which waves of failures have followed periods of monetary tightening.

\begin{figure}[!htbp]
    \centering
    \includegraphics[width=\linewidth]{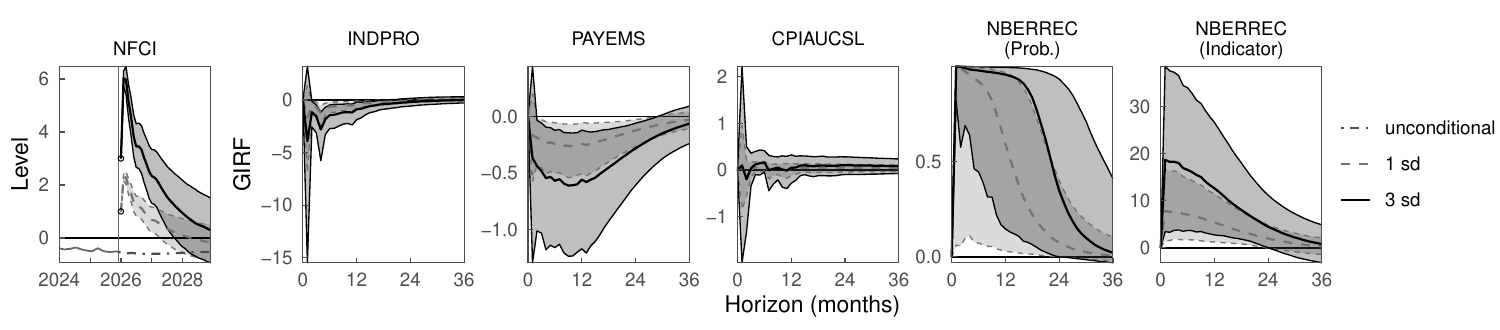}
    \caption{Growth-at-risk scenarios: \texttt{NFCI} is shocked on impact by $1$ and $3$ standard deviations.}
    \label{fig:cf_gar}
    \fnote{\textit{Notes:} The left panel shows the imposed impact restrictions (circles) alongside the unconditional forecast (dash-dotted); posterior medians with 68 percent credible sets. The remaining panels show GIRFs as in Figure~\ref{fig:cf_banks}; for \texttt{NBERREC} the responses of the recession probability (Prob.) and of the business cycle indicator (Indicator) are shown in separate panels. Figure~\ref{fig:cf_gar_tail} shows the corresponding 5/95 percentile bands.}
\end{figure}

\textit{Growth-at-risk.} Following the growth-at-risk literature \citep{adrian2019vulnerable}, which relates the lower tail of the distribution of future growth to financial conditions, we shock the \texttt{NFCI} on impact by one and three standard deviations. Since the index is standardized, these correspond to moderate and severe tightenings of financial conditions. Figure~\ref{fig:cf_gar} shows that financial conditions tighten further in the month after the shock and then ease gradually toward their unconditional path over the three-year horizon. Under the severe scenario, the growth rate of industrial production falls by about four percent on impact and remains below its unconditional path for about two years, so that the level remains permanently lower; employment growth declines for a year, to about $-0.5$ percent per month, and returns to its unconditional path after three years, while inflation hardly responds. The recession probability rises to nearly one for about a year and a half before it declines; under the moderate scenario it rises by a similar amount on impact but fades within about a year and a half. The responses of the continuous variables scale roughly with the size of the shock, whereas the recession probability does not, which reflects the nonlinearity of the measurement equation. Appendix~\ref{app:emp} complements the GIRFs by reporting the tails of the conditional distribution, where asymmetries associated with downside risks to the real economy would materialize.

\begin{figure}[!htbp]
    \centering
    \includegraphics[width=\linewidth]{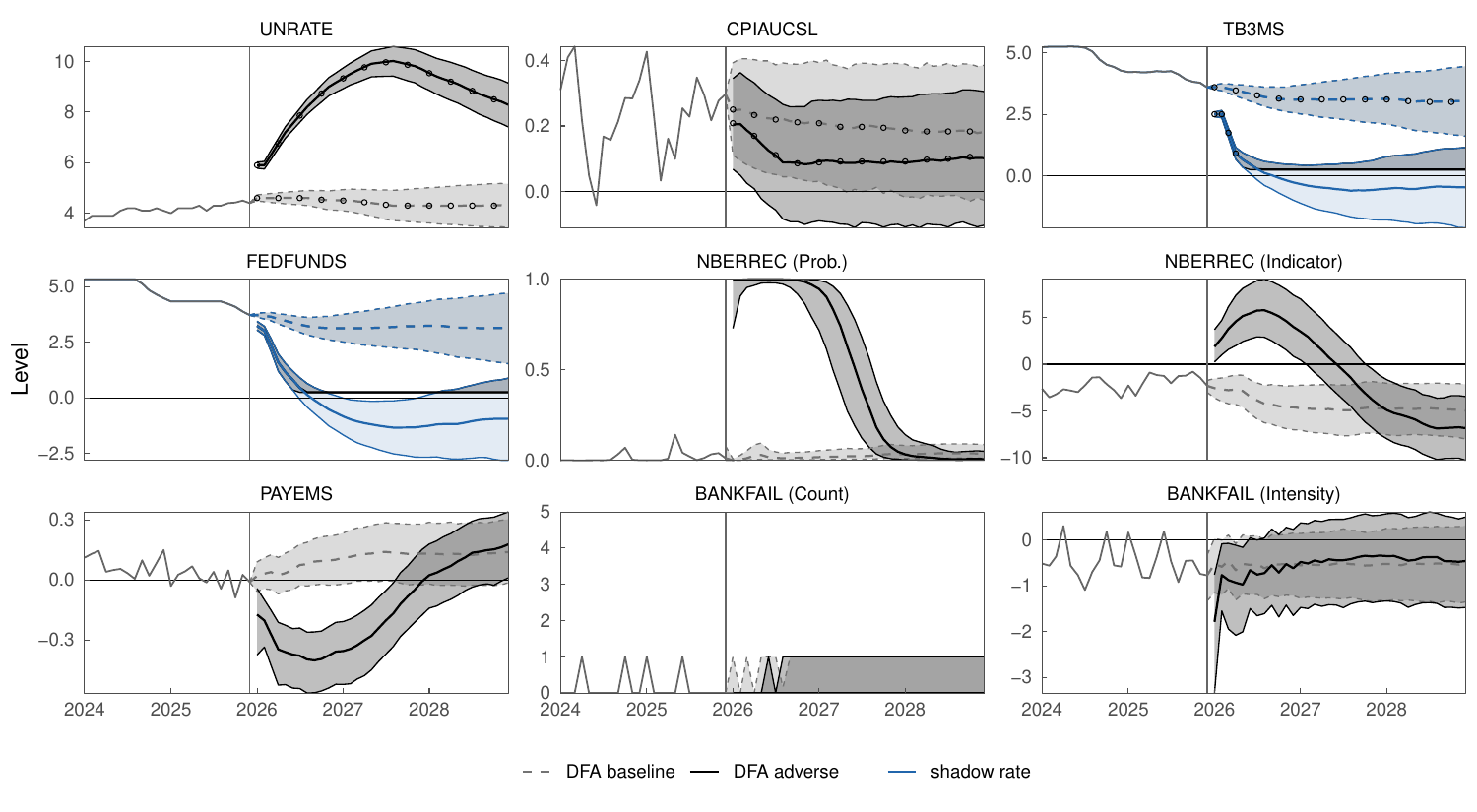}
    \caption{Stress test scenarios: conditional forecasts under the baseline and severely adverse supervisory scenarios of the 2026 Dodd-Frank Act stress test.}
    \label{fig:cf_dfa}
    \fnote{\textit{Notes:} Posterior medians alongside 68 percent credible sets of the conditional forecasts; the grey line shows the in-sample posterior median for the preceding two years, and shadow rates are shown in blue. Circles mark the imposed restrictions on \texttt{UNRATE}, \texttt{CPIAUCSL} and \texttt{TB3MS} (the restrictions on \texttt{GS5} and \texttt{GS10} are not shown); the remaining panels show the implied paths of the other variables.}
\end{figure}

\textit{Stress tests.} We impose the supervisory baseline and severely adverse scenarios of the 2026 Dodd-Frank Act stress test on the unemployment rate, CPI inflation, and the three-month, five-year and ten-year Treasury rates (\texttt{UNRATE}, \texttt{CPIAUCSL}, \texttt{TB3MS}, \texttt{GS5}, \texttt{GS10}). The scenario involves interest rates that are subject to censoring: where the supervisory path lies below the $0.25$ percent threshold, we leave the rate unrestricted, and the model reports its shadow rate. Figure~\ref{fig:cf_dfa} shows the conditional forecasts. Under the severely adverse scenario, the unemployment rate rises to ten percent by mid-2027, the three-month rate reaches the ELB in the third quarter of 2026, and the federal funds rate, on which the scenario imposes no restriction, follows it, with a shadow rate of about $-1.3$ percent in 2027. The recession probability is close to one throughout 2026 and falls to near zero by the end of 2027, and employment growth falls to about $-0.4$ percent per month before it recovers. Bank failures hardly respond within the horizon, in line with the lag of two to three years with which failures followed the downturns of the early 1980s and of 2008. Under the baseline scenario, the recession probability stays near zero.

\begin{figure}[!htbp]
    \centering
    \includegraphics[width=\linewidth]{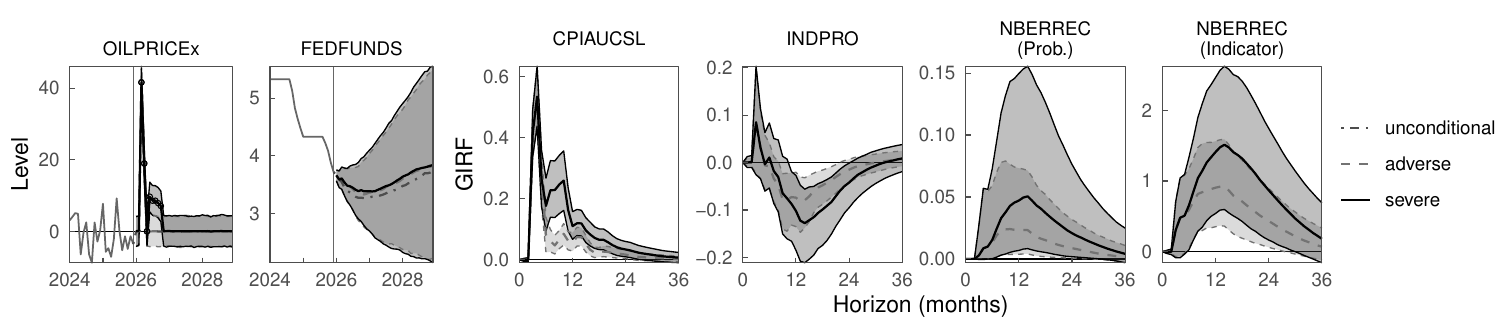}
    \caption{Oil price scenarios: adverse (one-quarter) and severe (three-quarter) oil price shock paths are imposed on \texttt{OILPRICEx}, following \citet{kilian2026iran}.}
    \label{fig:cf_oil}
    \fnote{\textit{Notes:} The two left panels show the imposed oil price paths (circles mark the restrictions) and the implied path of the federal funds rate alongside the unconditional forecast (dash-dotted); posterior medians with 68 percent credible sets. The remaining panels show GIRFs as in Figure~\ref{fig:cf_banks}.}
\end{figure}

\textit{Oil price scenarios.} We impose an adverse and a severe oil price path on \texttt{OILPRICEx}, taken from the scenarios that \citet{kilian2026iran} calibrate to the 2026 Iran war: the adverse path raises the oil price sharply for one quarter, the severe path for three quarters, with an initial increase of about 40 percent. Figure~\ref{fig:cf_oil} shows the imposed paths, the implied path of the federal funds rate, and the GIRFs of inflation, industrial production, the recession probability and the business cycle indicator. Under the severe scenario, monthly CPI inflation rises by up to 0.5 percentage points in the months of the shock and the effect dies out within two years; industrial production declines by about 0.1 percent after a year; and the recession probability rises by up to five percentage points. Under the adverse scenario, the initial response of inflation is the same but fades faster, and the responses of industrial production and of the recession probability are about half as large. The federal funds rate is essentially unchanged relative to its unconditional path under both scenarios.

The five exercises together illustrate the versatility of the framework. The questions they address would otherwise call for different models: a target for the recession probability is a question for a probit model, a wave of bank failures for a count model, a supervisory stress test with interest rates at the lower bound for a censored model, and financial and oil price shocks for a standard VAR. In the proposed framework each is answered within one model, so that a scenario can combine restrictions on variables of different types, as the stress test does, and the responses of all variable types, including the recession probability, the expected number of bank failures and the shadow rates, are obtained from the same predictive distribution.

\FloatBarrier
\section{Concluding Remarks and Future Research} \label{s:conclusion}

We have extended the VAR to a mix of binary, censored, count and unrestricted variables, with missing observations of any type handled in the same way. Each restricted observation is linked to a latent Gaussian variable, the counts through a threshold function from the Box--Cox family, and the latent variables are drawn, jointly or equation by equation, from a truncated normal distribution with a banded precision matrix. We have demonstrated the proposed framework in an application that delivers a business cycle indicator, the intensity of banking distress and shadow rates within one system, together with conditional forecasts in which the restrictions are placed on these variables themselves.

The framework can be extended in several directions. Ordinal variables, such as sovereign credit ratings, fit the same latent-variable representation with ordered thresholds \citep{zhang2015bayesian}, and the conditional forecasting machinery applies to them without change. In addition, the latent representation makes structural analysis available for restricted variables. Impulse responses of the recession probability, of the expected number of bank failures and of the shadow rates to identified shocks can be computed from the same posterior draws. The identification schemes developed for standard VARs, such as recursive or sign restrictions, carry over to the latent system. We leave these extensions for future research.


{\setstretch{1.20}
\small\addcontentsline{toc}{section}{References}
\bibliographystyle{custom.bst}
\bibliography{lit}}\normalsize\clearpage\doublespacing

\begin{appendices}
\begin{center}
    \huge\textsff{\textbf{Online Appendix}}
\end{center}

\setcounter{page}{1}
\setcounter{section}{0}
\setcounter{figure}{0}
\setcounter{table}{0}
\setcounter{equation}{0}
\setcounter{footnote}{0}

\renewcommand\thesection{\Alph{section}}
\renewcommand\theequation{\Alph{section}.\arabic{equation}}
\renewcommand\thefigure{\Alph{section}.\arabic{figure}}
\renewcommand\thetable{\Alph{section}.\arabic{table}}

\section{Joint Predictive Distribution} \label{app:pred}

This appendix derives the joint predictive distribution of the latent and observed variables over the forecast horizon, conditional on the parameters and the volatilities. It then shows how the conditioning assumptions of Section~\ref{sec:cond_fc} are imposed on this distribution, which shifts the means of the restricted elements and preserves the covariance of the unconditional forecast.

The model in \eqref{eq:VAR} can be written in structural form as
\begin{equation}
    \bm{B}_0\bm{y}_t = \bm{b} + \bm{B}_1\bm{y}_{t-1} + \cdots + \bm{B}_P\bm{y}_{t-P} + \bm{O}_t\bm{H}_t^{1/2}\bm{\epsilon}_t,\label{eq:SVAR}
\end{equation}
where $\bm{B}_0 = \bm{A}_0$, $\bm{B}_p = \bm{A}_0\bm{A}_p$ for $p = 1, \hdots, P$ and $\bm{b} = \bm{A}_0\bm{a}$. Following \citet{chan2024conditional}, we stack \eqref{eq:SVAR} over the forecast horizon $t = T+1, \hdots, T+F$. Let $\bm{y}_f = (\bm{y}_{T+1}', \hdots, \bm{y}_{T+F}')'$ and define
\begin{equation*}
    \widetilde{\bm{M}}_f = \begin{bmatrix}
        \bm{B}_0      & \bm{0}_{n\times n} & \cdots      & \cdots      & \cdots      & \cdots      & \bm{0}_{n\times n}\\
        -\bm{B}_1     & \bm{B}_0           & \bm{0}_{n\times n} & \cdots & \cdots      & \cdots      & \bm{0}_{n\times n}\\
        -\bm{B}_2     & -\bm{B}_1          & \bm{B}_0    & \bm{0}_{n\times n} & \cdots & \cdots      & \bm{0}_{n\times n}\\
        \vdots        & \ddots             & \ddots      & \ddots      & \ddots      &             & \vdots     \\
        -\bm{B}_P     & \cdots             &             & -\bm{B}_1  & \bm{B}_0    & \bm{0}_{n\times n} & \vdots \\
        \bm{0}_{n\times n} & \ddots        &             & \ddots      & \ddots      & \ddots      & \vdots \\
        \vdots        & \ddots             & \ddots      &             & \ddots      & \ddots      & \vdots \\
        \bm{0}_{n\times n} & \cdots        & \bm{0}_{n\times n} & -\bm{B}_P & \cdots & -\bm{B}_1  & \bm{B}_0
    \end{bmatrix}, \quad
    \widetilde{\bm{m}}_f = \begin{bmatrix}
        \bm{b} + \sum_{p=1}^{P}\bm{B}_p\bm{y}_{T+1-p} \\
        \bm{b} + \sum_{p=2}^{P}\bm{B}_p\bm{y}_{T+2-p} \\
        \bm{b} + \sum_{p=3}^{P}\bm{B}_p\bm{y}_{T+3-p} \\
        \vdots \\
        \bm{b} + \bm{B}_P\bm{y}_T \\
        \bm{b} \\
        \vdots \\
        \bm{b}
    \end{bmatrix},
\end{equation*}
and $\bm{\Omega}_f^{1/2} = \bdiag(\bm{O}_{T+1}\bm{H}_{T+1}^{1/2}, \hdots, \bm{O}_{T+F}\bm{H}_{T+F}^{1/2})$, where the log-volatilities over the forecast horizon are simulated forward from their state equations and the outlier factors are set to one. The stacked system is $\widetilde{\bm{M}}_f\bm{y}_f = \widetilde{\bm{m}}_f + \bm{\Omega}_f^{1/2}\bm{\epsilon}_f$ with $\bm{\epsilon}_f \sim \mathcal{N}(\bm{0}_{nF}, \bm{I}_{nF})$. With $\bm{M}_f = \bm{\Omega}_f^{-1/2}\widetilde{\bm{M}}_f$ and $\bm{m}_f = \bm{\Omega}_f^{-1/2}\widetilde{\bm{m}}_f$, it becomes $\bm{M}_f\bm{y}_f = \bm{m}_f + \bm{\epsilon}_f$, so that $\bm{y}_f = \bm{M}_f^{-1}\bm{m}_f + \bm{M}_f^{-1}\bm{\epsilon}_f$ and, conditional on the parameters and on the volatilities over the forecast horizon,
\begin{equation}
    \bm{y}_f \sim \mathcal{N}\left(\bm{\mu}_f, \bm{\Sigma}_f\right), \quad \bm{\mu}_f = \bm{M}_f^{-1}\bm{m}_f, \quad \bm{\Sigma}_f = (\bm{M}_f'\bm{M}_f)^{-1}. \label{eq:pred_dist}
\end{equation}
Draws from \eqref{eq:pred_dist} are obtained by drawing $\bm{\epsilon}_f$ and solving the block lower triangular system $\bm{M}_f\bm{y}_f = \bm{m}_f + \bm{\epsilon}_f$ by forward substitution, and the unconditional forecasts collect these draws across the posterior draws of the parameters and the simulated volatilities. The draws of the latent binary, count and censored variables are mapped to the observed outcomes through the measurement equations \eqref{eq:linkbinary} to \eqref{eq:linkcount}, so that the predictive distribution of an observed variable is the distribution of these mapped draws. For the homoskedastic version of the model, in which $\bm{\Sigma}$ is drawn directly, we set $\bm{B}_0$ to the inverse of the lower Cholesky factor of $\bm{\Sigma}$, which satisfies $\bm{B}_0\bm{\Sigma}\bm{B}_0' = \bm{I}_n$, and $\bm{\Omega}_f^{1/2} = \bm{I}_{nF}$; then $\bm{M}_f = \widetilde{\bm{M}}_f$ and $\bm{m}_f = \widetilde{\bm{m}}_f$.

A scenario restricts $q$ elements or linear combinations of $\bm{y}_f$, observed or latent, through a $q \times nF$ matrix $\bm{R}$. We impose the restriction $\bm{R}\bm{y}_f \sim \mathcal{N}(\bm{r}, \bm{W})$, where $\bm{r}$ collects the imposed values and $\bm{W}$ is the restriction covariance; hard restrictions correspond to $\bm{W} = \bm{0}$. With $\bm{Q} = \bm{R}\bm{\Sigma}_f\bm{R}'$, the conditional forecast distribution is Gaussian with mean $\bm{\mu}_f + \bm{\Sigma}_f\bm{R}'\bm{Q}^{-1}(\bm{r} - \bm{R}\bm{\mu}_f)$ and covariance $\bm{\Sigma}_f - \bm{\Sigma}_f\bm{R}'\bm{Q}^{-1}\bm{R}\bm{\Sigma}_f + \bm{\Sigma}_f\bm{R}'\bm{Q}^{-1}\bm{W}\bm{Q}^{-1}\bm{R}\bm{\Sigma}_f$. In Section~\ref{sec:cond_fc} we set $\bm{W} = \bm{Q}$, so that the restricted elements have mean $\bm{r}$ and the covariance of the conditional forecast equals $\bm{\Sigma}_f$: the restrictions shift the means and preserve the uncertainty of the unconditional forecast. The conditional distribution depends on the reduced-form objects $\bm{\mu}_f$ and $\bm{\Sigma}_f$ only, so no structural identification is needed. For computation, we write $\bm{y}_f = \bm{\mu}_f + \bm{M}_f^{-1}\bm{\epsilon}_f$ and obtain the conditional draw by adjusting the draw of $\bm{\epsilon}_f$ used for the unconditional path \citep[see][]{antolin2021structural}, so that the difference between the conditional and the unconditional path of each draw is the response to the restrictions; the GIRFs in Section~\ref{sec:cond_fc} are the posterior distributions of these differences. A restriction on a count is imposed at the logarithm of the midpoint of the interval of the restricted count, and a restriction on the recession probability at the corresponding quantile of the business cycle indicator, which targets the probability exactly when the innovation variance of the indicator equals one and approximately under stochastic volatility. In both cases the restriction fixes the mean of the latent variable, and the realized count or binary outcome remains random under the scenario.

\section{Data and Additional Empirical Results} \label{app:emp}

\begin{figure}[!htbp]
    \centering
    \includegraphics[width=\linewidth]{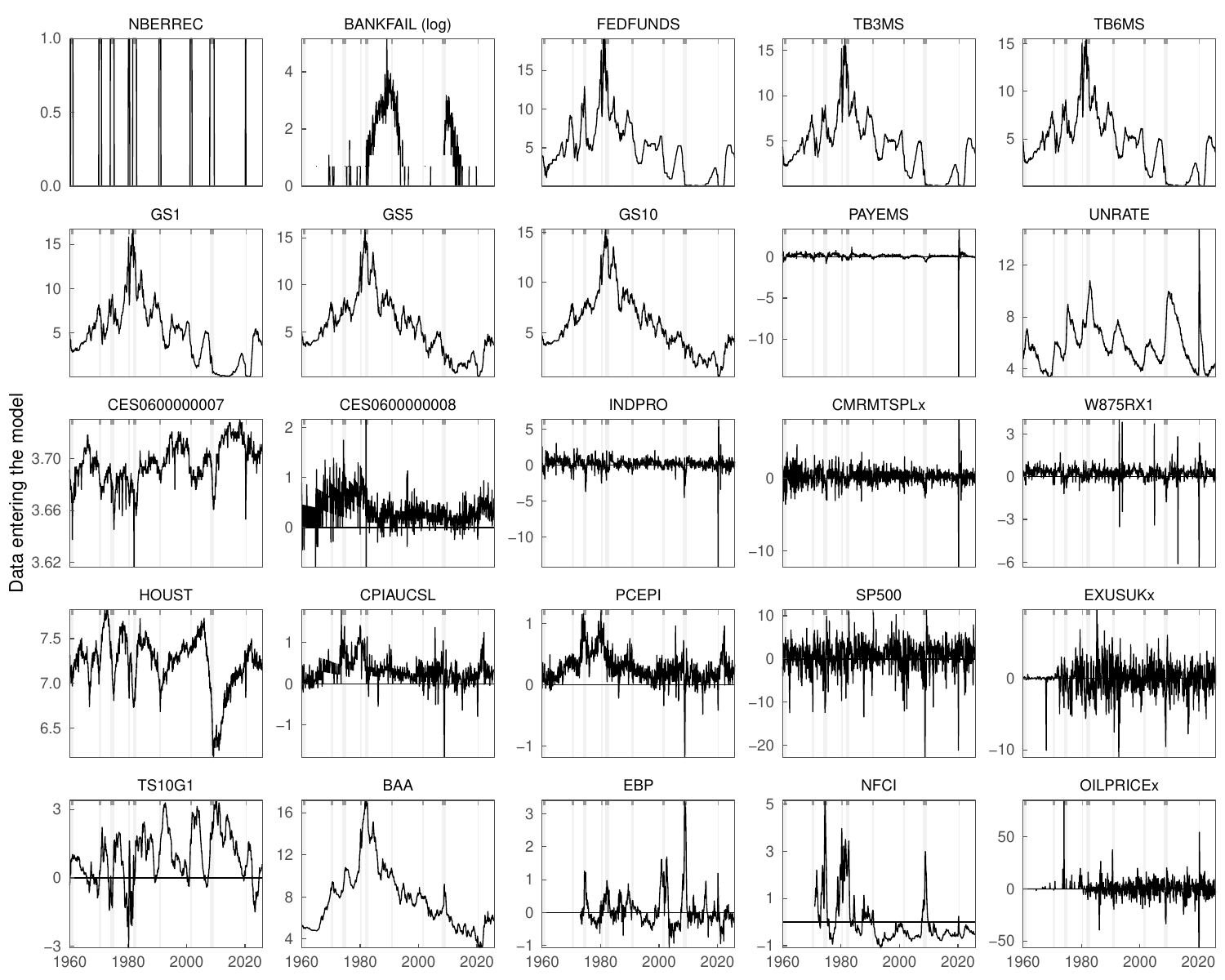}
    \caption{Time series of the 25 variables entering the model, 1960:01 to 2025:12.}
    \label{fig:data}
    \fnote{\textit{Notes:} Monthly observations from 1960:01 to 2025:12, in the units in which they enter the VAR, ordered by type: the binary, the count, the censored, and the unrestricted variables. Table~\ref{tab:data} lists definitions and transformations. The count variable is plotted as the logarithm of the count, which is the threshold scale of its measurement equation; months without bank failures have no logarithm and are left blank, and they are not missing observations. The remaining gaps mark the observations that the model treats as missing, \texttt{EBP} before 1973:01 and \texttt{NFCI} before 1971:01.}
\end{figure}

\begin{figure}[!htbp]
    \centering
    \includegraphics[width=\linewidth]{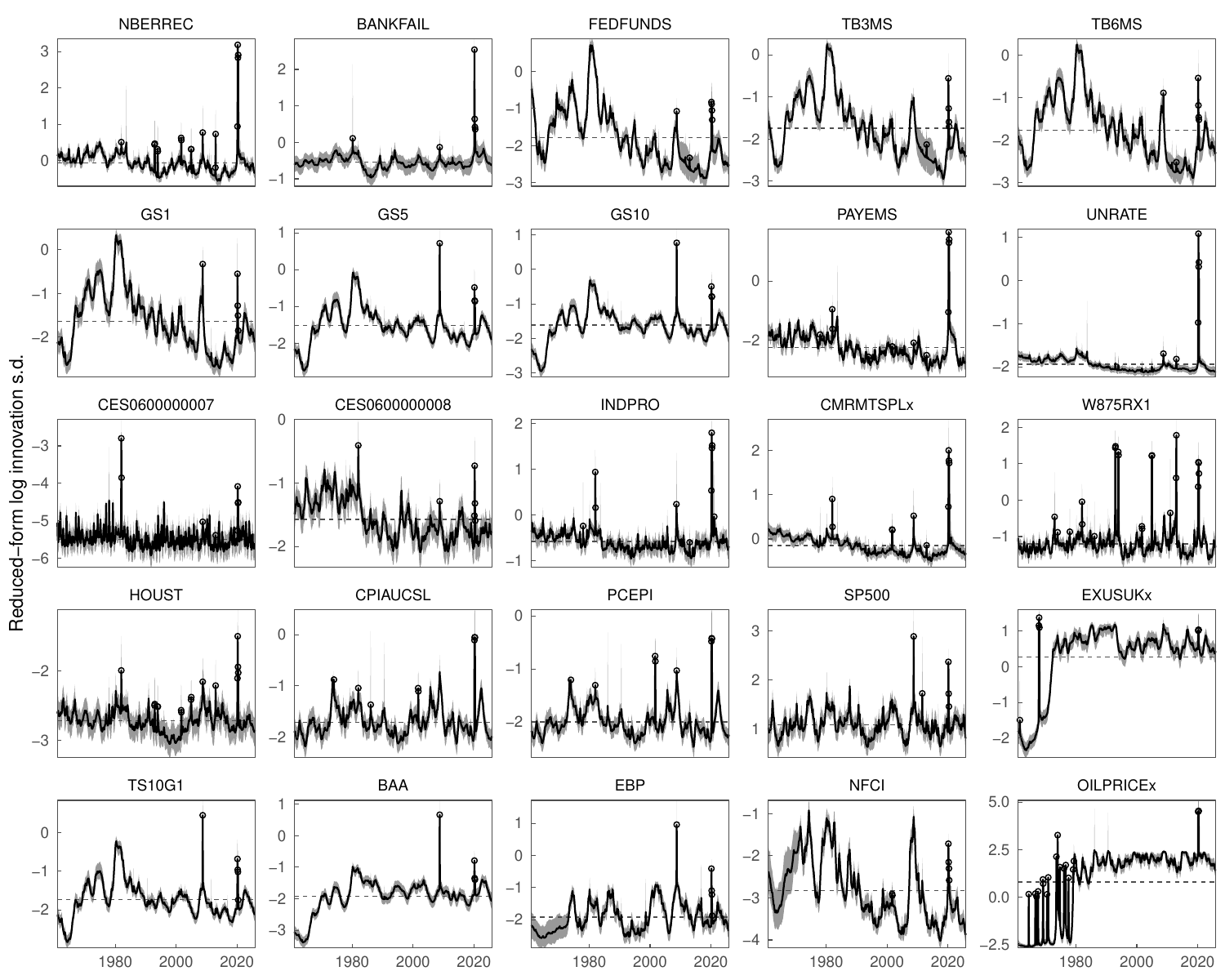}
    \caption{Estimates of the reduced-form log innovation standard deviations, $\log\sqrt{[\bm{\Sigma}_t]_{ii}}$, for all variables, on the scale of the latent variable for the binary, count and censored variables and on the scale of the observed data otherwise.}
    \label{fig:sigma_all}
    \fnote{\textit{Notes:} See Figure~\ref{fig:latent_sigma}, which shows the first three variables of this figure for the pre-COVID-19 and the COVID-19 sample separately. Here, each panel covers the full sample and the vertical scale is not restricted to the range of the posterior median.}
\end{figure}

\begin{figure}[!htbp]
    \centering
    \includegraphics[width=\linewidth]{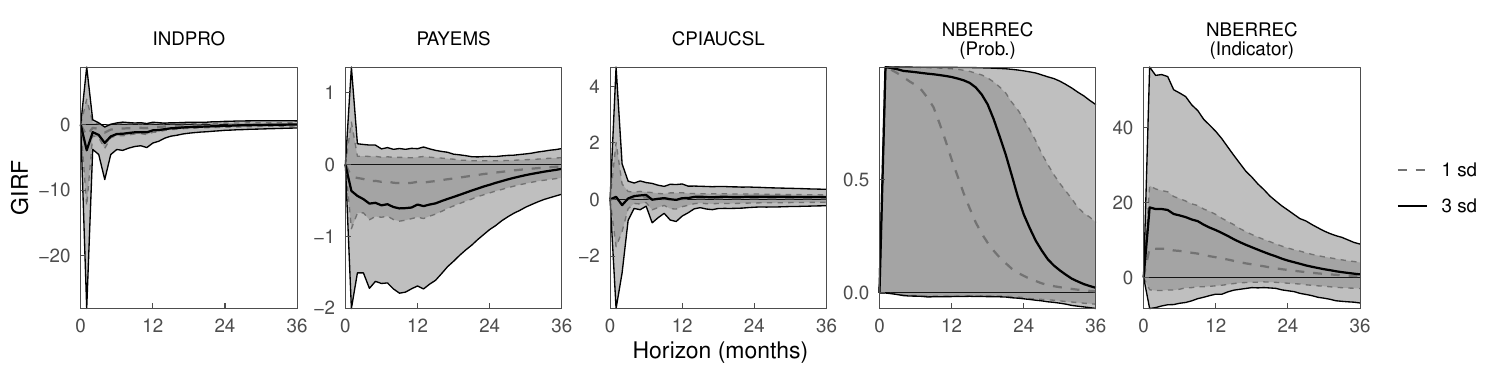}
    \caption{Growth-at-risk scenarios: 5/95 percentile bands of the GIRFs shown in Figure~\ref{fig:cf_gar}.}
    \label{fig:cf_gar_tail}
    \fnote{\textit{Notes:} Posterior medians alongside 90 percent credible sets of the draw-wise difference between the conditional and the unconditional predictive distribution.}
\end{figure}

\begin{figure}[!htbp]
    \centering
    \includegraphics[width=\linewidth]{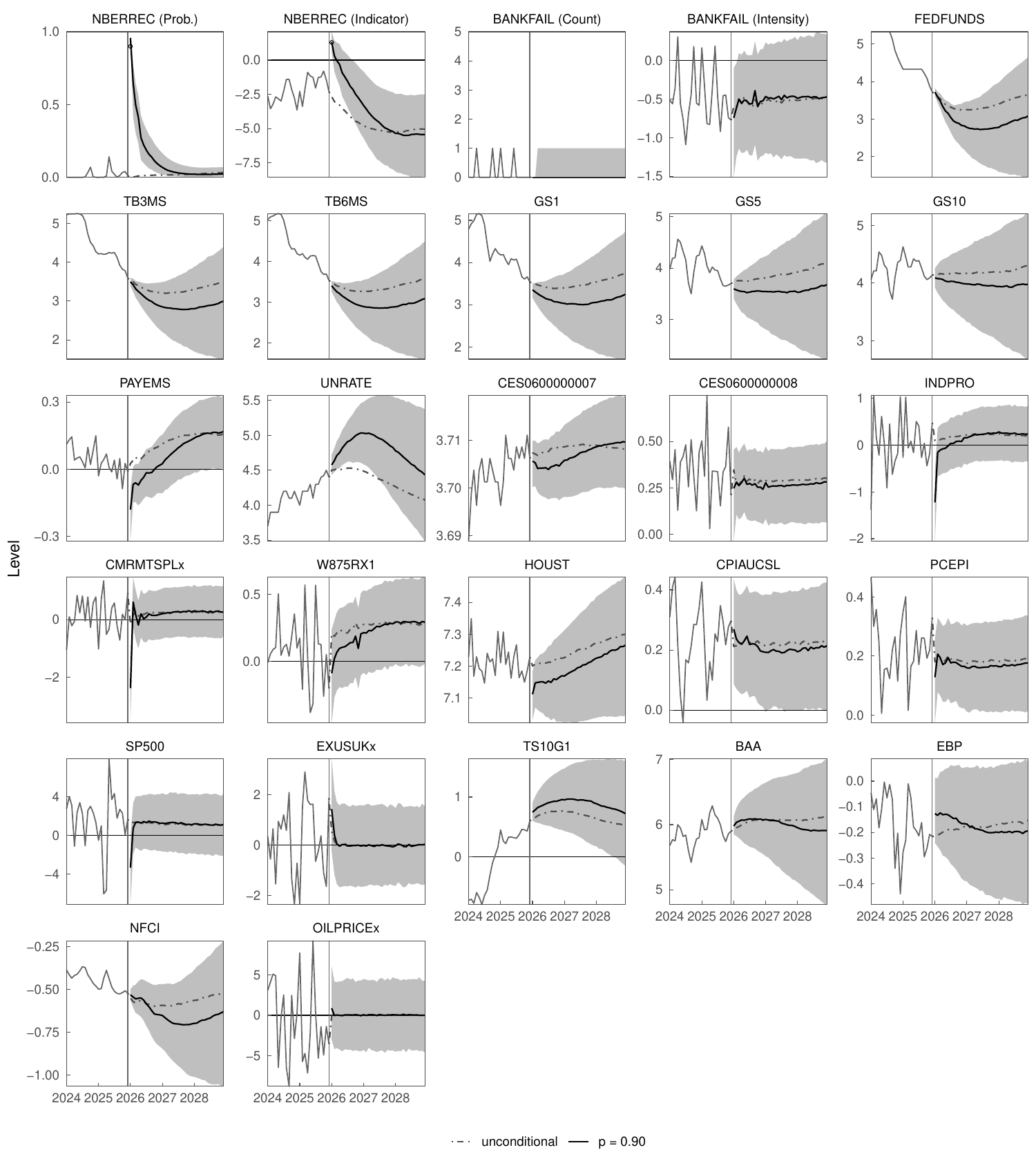}
    \caption{Recession probability scenario ($p = 0.9$): conditional forecasts for all variables.}
    \label{fig:cf_recession_all}
    \fnote{\textit{Notes:} Posterior medians of the conditional (black solid) and unconditional (grey dash-dotted) forecasts, with 68 percent credible sets for the conditional forecast; the preceding two years show the in-sample posterior median. Circles mark the imposed restrictions. Panels labeled ``(Indicator)'', ``(Intensity)'' and ``(Shadow rate)'' show the associated latent states on their model scale.}
\end{figure}

\begin{figure}[!htbp]
    \centering
    \includegraphics[width=\linewidth]{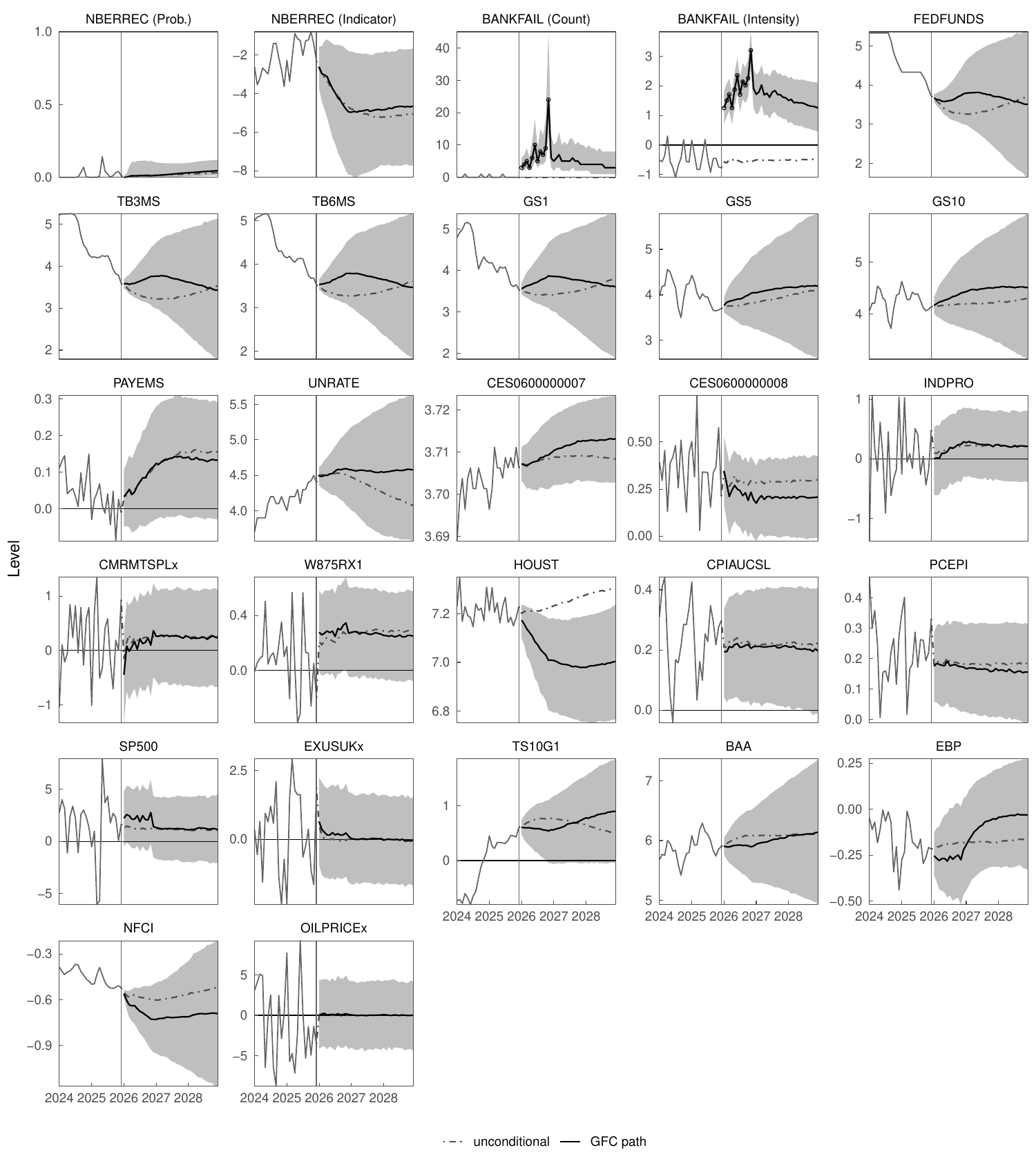}
    \caption{Bank failure scenario (Global Financial Crisis path): conditional forecasts for all variables.}
    \label{fig:cf_banks_all}
    \fnote{\textit{Notes:} See Figure~\ref{fig:cf_recession_all}.}
\end{figure}

\begin{figure}[!htbp]
    \centering
    \includegraphics[width=\linewidth]{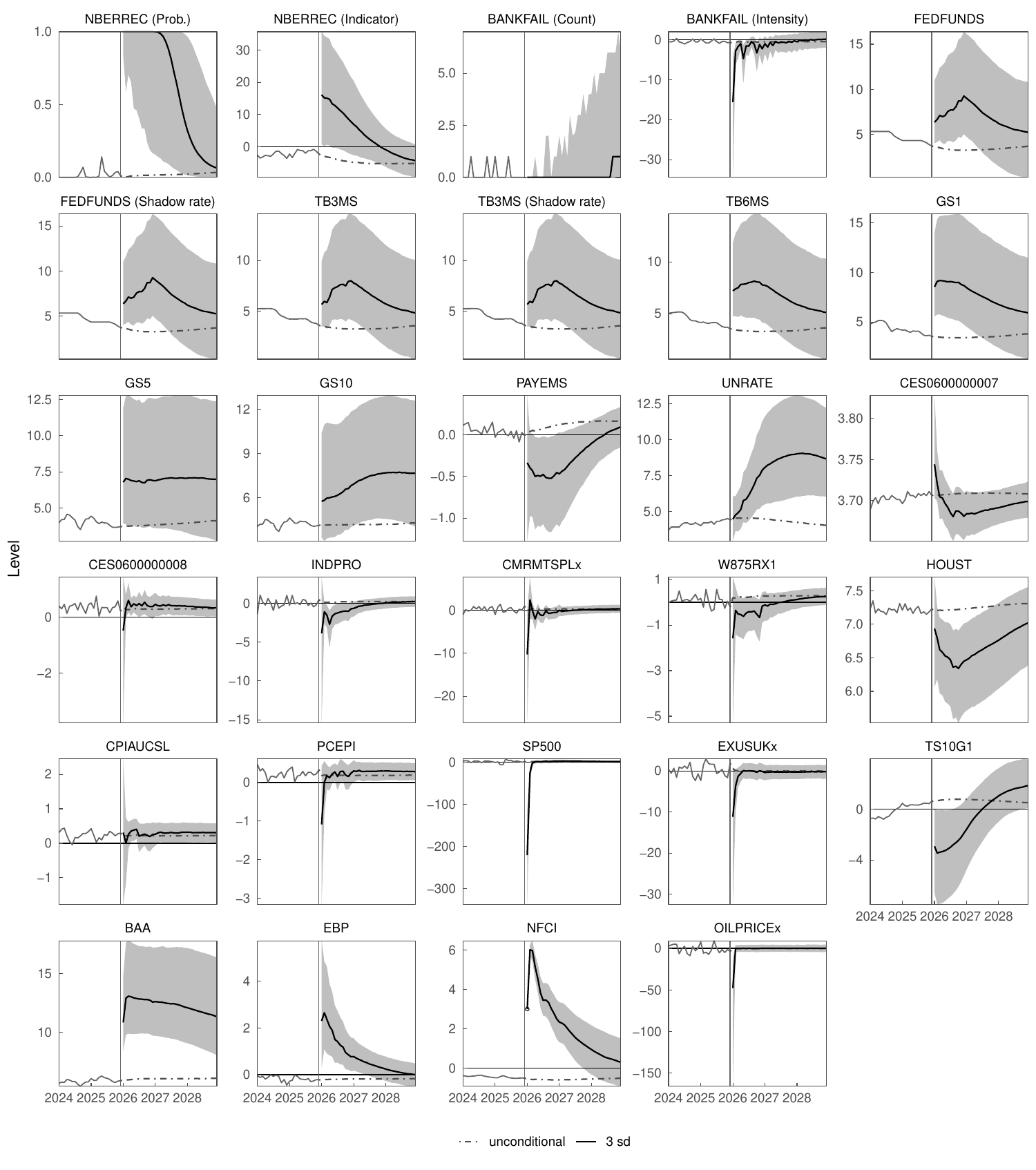}
    \caption{Growth-at-risk scenario ($3$ standard deviations): conditional forecasts for all variables.}
    \label{fig:cf_gar_all}
    \fnote{\textit{Notes:} See Figure~\ref{fig:cf_recession_all}.}
\end{figure}

\begin{figure}[!htbp]
    \centering
    \includegraphics[width=\linewidth]{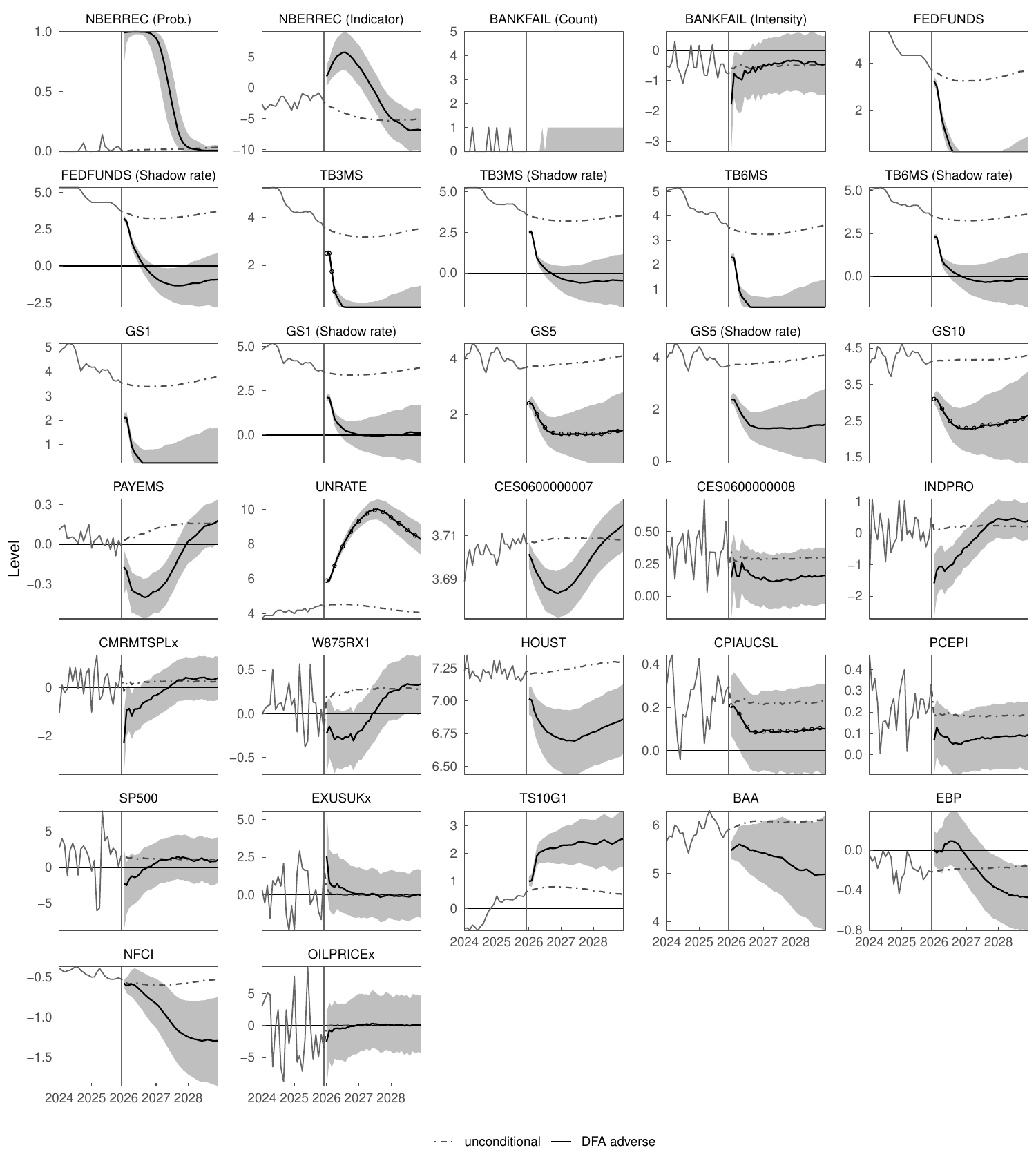}
    \caption{Stress test scenario (severely adverse): conditional forecasts for all variables.}
    \label{fig:cf_dfa_all}
    \fnote{\textit{Notes:} See Figure~\ref{fig:cf_recession_all}.}
\end{figure}

\begin{figure}[!htbp]
    \centering
    \includegraphics[width=\linewidth]{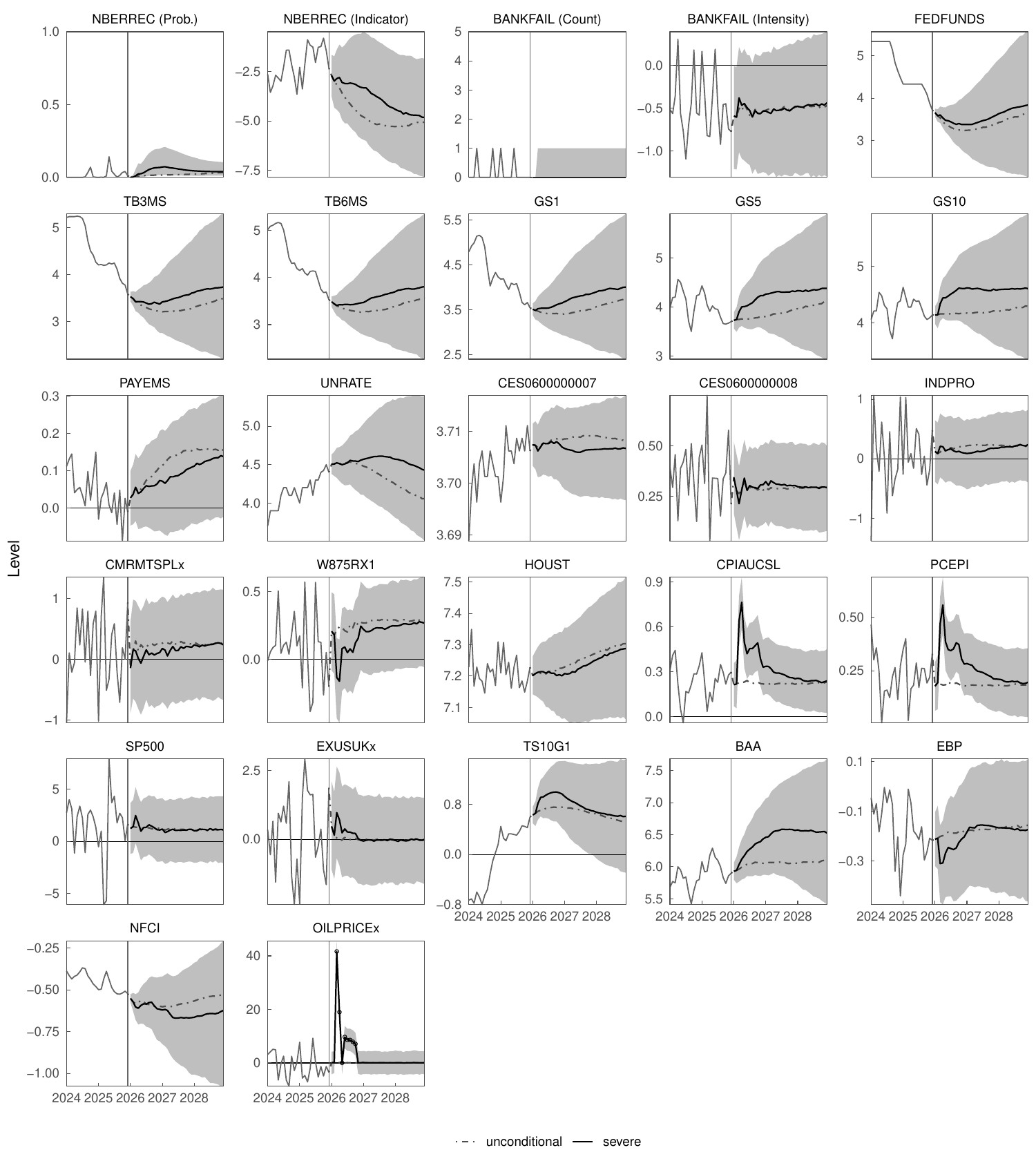}
    \caption{Oil price scenario (severe): conditional forecasts for all variables.}
    \label{fig:cf_oil_all}
    \fnote{\textit{Notes:} See Figure~\ref{fig:cf_recession_all}.}
\end{figure}

This appendix lists the variables entering the model (Table~\ref{tab:data}), plots them (Figure~\ref{fig:data}), and reports the reduced-form log innovation standard deviations for the full set of variables (Figure~\ref{fig:sigma_all}). It then repeats the estimates of the latent states of Section~\ref{sec:latent_states} for the homoskedastic version of the model. This version replaces stochastic volatility and the outlier component by a constant covariance matrix $\bm{\Sigma}_t = \bm{\Sigma}$ with the inverse Wishart prior $\bm{\Sigma}\sim\mathcal{W}^{-1}(n+2,\bm{I}_n)$; all other specification choices are unchanged. Each draw of $\bm{\Sigma}$ is normalized as $\bm{\Sigma}^{\ast} = \bm{D}^{-1}\bm{\Sigma}\bm{D}^{-1}$ with $d_i = \sqrt{[\bm{\Sigma}]_{ii}}$ for the binary variables and $d_i = 1$ otherwise, which is the normalization of \citet{dueker2005dynamic}. Finally, it reports additional results from the conditional forecasting exercises of Section~\ref{sec:cond_fc}: Figure~\ref{fig:cf_gar_tail} shows the 5/95 percentile bands of the growth-at-risk GIRFs, and Figures~\ref{fig:cf_recession_all} to \ref{fig:cf_oil_all} show, for the most severe variant of each scenario family, the conditional forecasts of all variables and their latent states.

\begin{table}[!htbp]
\centering\footnotesize
\caption{The 25 monthly US macro-financial variables used in the application together with their transformations.}\label{tab:data}
\begin{threeparttable}
\begin{tabular*}{\textwidth}{@{\extracolsep{\fill}}llcc@{}}
\toprule
Mnemonic & Description & Type & Tfm. \\ \midrule
\texttt{NBERREC} & NBER recession indicator & \texttt{b} & -- \\
\texttt{BANKFAIL} & Number of FDIC-insured bank failures & \texttt{z} & --\tnote{a} \\
\texttt{FEDFUNDS} & Effective federal funds rate & \texttt{c} & -- \\
\texttt{TB3MS} & 3-month Treasury bill rate & \texttt{c} & -- \\
\texttt{TB6MS} & 6-month Treasury bill rate & \texttt{c} & -- \\
\texttt{GS1} & 1-year Treasury rate & \texttt{c} & -- \\
\texttt{GS5} & 5-year Treasury rate & \texttt{c} & -- \\
\texttt{GS10} & 10-year Treasury rate & \texttt{c} & -- \\
\texttt{PAYEMS} & All employees: total nonfarm & \texttt{u} & $100\,\Delta\log$ \\
\texttt{INDPRO} & Industrial production index & \texttt{u} & $100\,\Delta\log$ \\
\texttt{CMRMTSPLx} & Real manufacturing and trade industries sales & \texttt{u} & $100\,\Delta\log$ \\
\texttt{W875RX1} & Real personal income ex transfer receipts & \texttt{u} & $100\,\Delta\log$ \\
\texttt{UNRATE} & Civilian unemployment rate & \texttt{u} & -- \\
\texttt{CES0600000007} & Average weekly hours: goods-producing & \texttt{u} & $\log$ \\
\texttt{CES0600000008} & Average hourly earnings: goods-producing & \texttt{u} & $100\,\Delta\log$ \\
\texttt{PCEPI} & Personal consumption expenditures price index & \texttt{u} & $100\,\Delta\log$ \\
\texttt{CPIAUCSL} & Consumer price index: all items & \texttt{u} & $100\,\Delta\log$ \\
\texttt{HOUST} & Housing starts: total new privately owned & \texttt{u} & $\log$ \\
\texttt{SP500} & S\&P 500 composite stock price index & \texttt{u} & $100\,\Delta\log$ \\
\texttt{EXUSUKx} & U.S.--U.K. foreign exchange rate & \texttt{u} & $100\,\Delta\log$ \\
\texttt{OILPRICEx} & Crude oil price (spliced WTI and Cushing) & \texttt{u} & $100\,\Delta\log$ \\
\texttt{BAA} & Moody's seasoned Baa corporate bond yield & \texttt{u} & -- \\
\texttt{TS10G1} & Term spread: 10-year minus 1-year Treasury rate & \texttt{u} & -- \\
\texttt{EBP}\tnote{b} & Excess bond premium & \texttt{u} & -- \\
\texttt{NFCI}\tnote{c} & Chicago Fed national financial conditions index & \texttt{u} & -- \\
\bottomrule
\end{tabular*}
\begin{tablenotes}[flushleft]\footnotesize
\item \textit{Notes:} Monthly data; the estimation sample runs from 1960-01 to 2025-12. Type indicates the variable type: binary (\texttt{b}), count (\texttt{z}), censored (\texttt{c}), or unrestricted (\texttt{u}). Tfm.\ denotes the transformation applied to the raw series: none (--), natural logarithm ($\log$), or log-difference in percent ($100\,\Delta\log$). Sources: FRED-MD (2026-06 vintage), NBER business cycle chronology, \citet{gilchrist2012credit} for the excess bond premium (Federal Reserve Board update), Federal Reserve Bank of Chicago (NFCI), and the FDIC failures and assistance database (\texttt{BANKFAIL}). \textsuperscript{a} Enters as an integer count; the count measurement equation uses the log thresholds ($\psi_i = 0$). \textsuperscript{b} Available from 1973-01 onwards. \textsuperscript{c} Available from 1971-01 onwards.
\end{tablenotes}
\end{threeparttable}
\end{table}

\begin{figure}[!htbp]
    \centering
    \includegraphics[width=\linewidth]{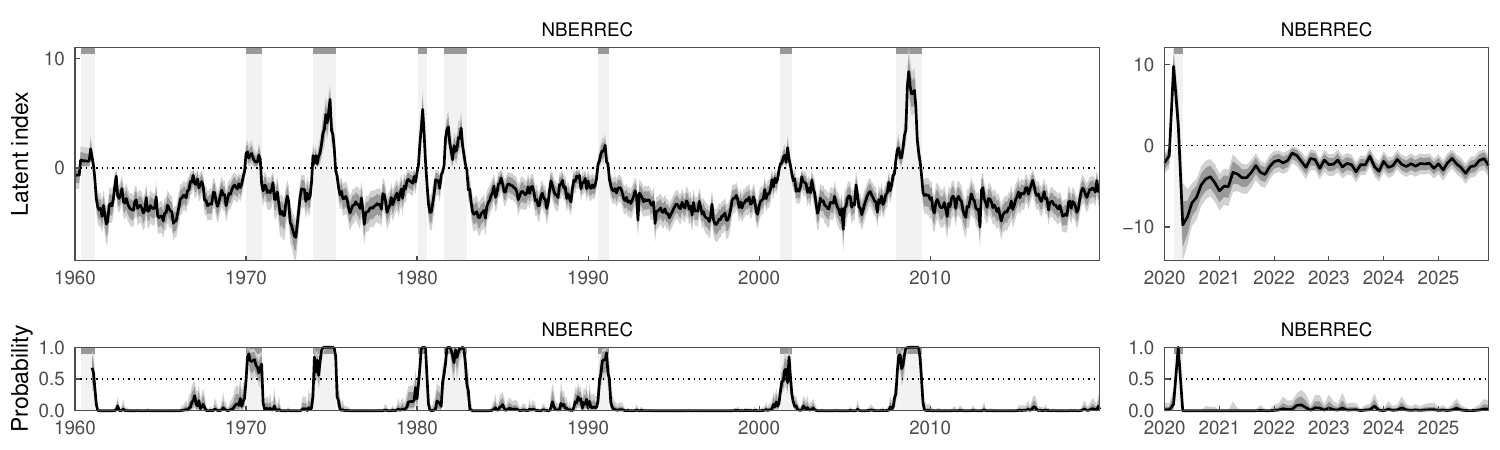}
    \caption{Homoskedastic model: estimates of the business cycle indicator underlying \texttt{NBERREC} (upper panels) and the implied recession probabilities (lower panels).}
    \label{fig:latent_binary_hom}
    \fnote{\textit{Notes:} See Figure~\ref{fig:latent_binary}.}
\end{figure}

\begin{figure}[!htbp]
    \centering
    \includegraphics[width=\linewidth]{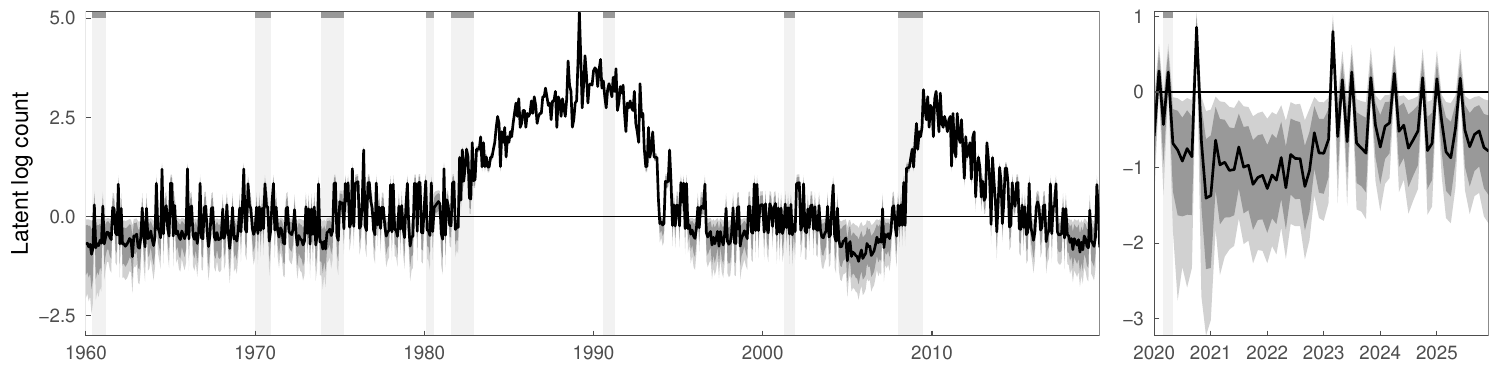}
    \caption{Homoskedastic model: estimates of the intensity of banking distress underlying the number of FDIC-insured bank failures (\texttt{BANKFAIL}).}
    \label{fig:latent_count_hom}
    \fnote{\textit{Notes:} See Figure~\ref{fig:latent_count}.}
\end{figure}

\begin{figure}[!htbp]
    \centering
    \includegraphics[width=\linewidth]{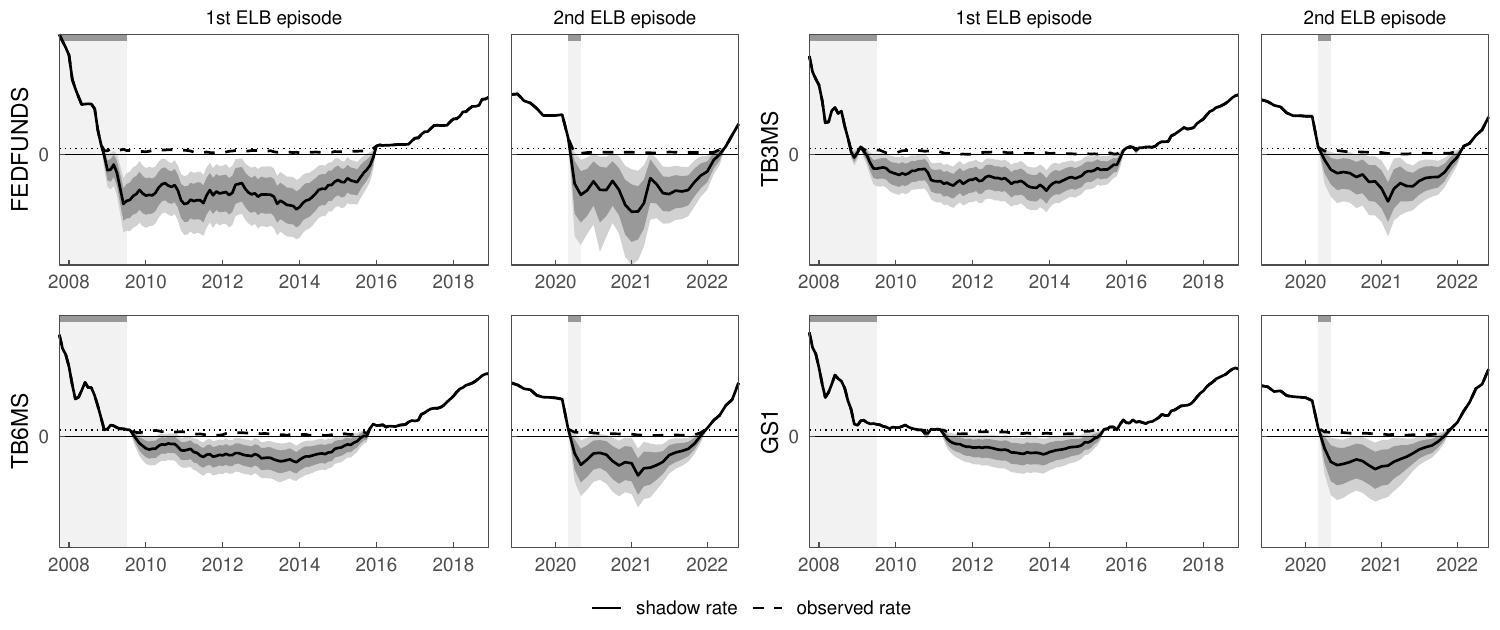}
    \caption{Homoskedastic model: estimates of the shadow rates underlying the interest rates subject to censoring at the ELB.}
    \label{fig:latent_censored_hom}
    \fnote{\textit{Notes:} See Figure~\ref{fig:latent_censored}.}
\end{figure}

\begin{figure}[!htbp]
    \centering
    \includegraphics[width=\linewidth]{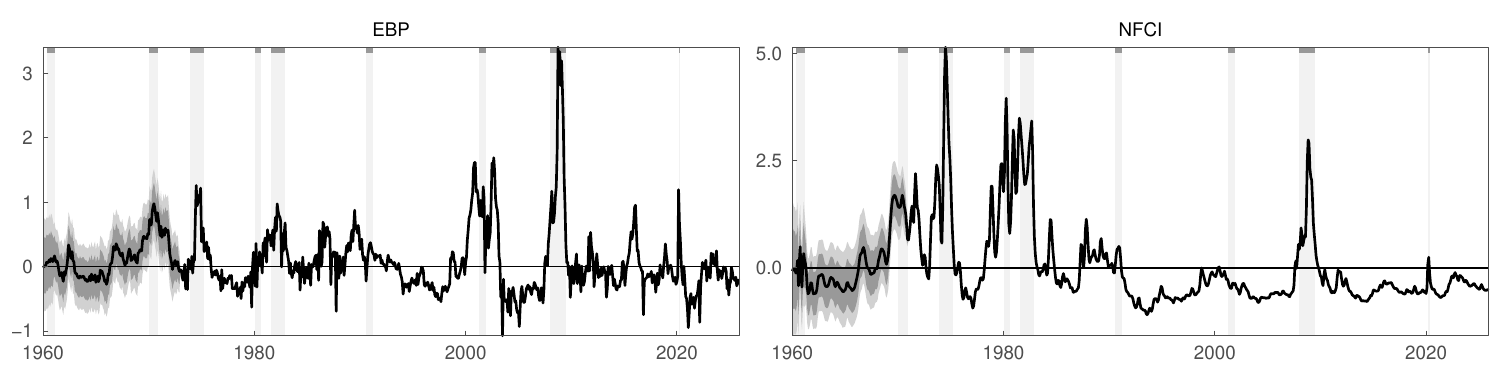}
    \caption{Homoskedastic model: imputed early observations of the excess bond premium (\texttt{EBP}) and the Chicago Fed national financial conditions index (\texttt{NFCI}), which are unavailable at the beginning of the sample.}
    \label{fig:latent_missing_hom}
    \fnote{\textit{Notes:} See Figure~\ref{fig:latent_missing}.}
\end{figure}

\FloatBarrier
\end{appendices}

\end{document}